\documentclass[11pt,a4paper]{article} 
\usepackage{jcappubNOHYP}
\usepackage{latexsym,amsmath,amssymb}
\usepackage{graphicx}
\usepackage{slashed}
\usepackage{bm}
\usepackage{epsfig}
\usepackage{mathrsfs}

\newcommand{\nn}{\nonumber}
\newcommand{\be}{\begin{eqnarray}}
\newcommand{\ee}{\end{eqnarray}}
\newcommand{\rar}{\rightarrow}

\def\dd{\text{d} }
\def\d{\partial}
\def\+{\dagger}
\def\<{\langle}
\def\>{\rangle}

\newcommand{\cph}{\varphi}

\newcommand{\cep}{\varepsilon}
\newcommand{\cH}{{\cal H}}
\newcommand{\cA}{{\cal A}}

\newcommand{\cL}{{\cal L}}

\newcommand{\cB}{{\cal B}}
\newcommand{\cS}{{\cal S}}
\newcommand{\cW}{{\cal W}}
\newcommand{\cT}{{\cal T}}
\newcommand{\cI}{{\cal I}}
\newcommand{\cO}{{\cal O}}

\newcommand{\Mpc}{\text{Mpc}}

\newcommand{\eb}{{\rm em}}

\newcommand{\pisubs}{{\Pi_S}}
\newcommand{\pisubsp}{{\Pi'_S}}
\newcommand{\ompi}{{\Omega^{-}_\Pi}}
\newcommand{\ompip}{{\Omega^{+}_\Pi}}
\newcommand{\omrho}{{\Omega^{-}_\rho}}

\newcommand{\source}{{S_{\rm em}}}

\newcommand{\ba}{\begin{eqnarray}}
\newcommand{\ea}{\end{eqnarray}}
\newcommand{\lag}{\mathcal{L}}
\newcommand{\bk}{{\mathbf k}}
\newcommand{\bp}{{\mathbf p}}
\newcommand{\bq}{{\mathbf q}}
\newcommand{\br}{{\mathbf r}}
\newcommand{\bs}{{\mathbf s}}

\newcommand{\bx}{{\mathbf x}}

\newcommand{\lp}{\left(}
\newcommand{\rp}{\right)}

\newcommand{\ep}{\epsilon}

\newcommand{\si}{\sigma}

\newcommand{\B}{\tilde{B}}

\title{Perturbations and non-Gaussianities in three-form inflationary magnetogenesis}

\author[a, b]{Federico~R.~Urban}
\author[b]{and Tomi~K.~Koivisto}

\affiliation[a]{Service de Physique Th\'eorique, Universit\'e ́Libre de Bruxelles, CP225, Boulevard du Triomphe, B-1050 Brussels, Belgium}
\affiliation[b]{Institute for Theoretical Astrophysics, University of Oslo, P.O.\ Box 1029 Blindern, N-0315 Oslo, Norway}

\date{\today}

\abstract{We reconsider magnetogenesis in the context of three-form inflation, and its backreaction.  In particular, we focus on first order perturbation theory during inflation and subsequent radiation era: we discuss the consistency of the perturbative approach, and elaborate on the possible non-Gaussian signatures of the model.}

\arxivnumber{}

\begin{document}
\maketitle

\section{Generalities}

Inflationary magnetogenesis has long been recognised as one of the most promising frameworks for the generation of today's observed cosmological magnetic fields, see~\cite{Han:2002ns,Subramanian:2008tt,Brown:2006wv,Kandus:2010nw,Widrow:2011hs}, and~\cite{Prokopec:2004au,Govoni:2004as,Bamba:2006ga,Beck:2008ty,Neronov:1900zz,Tavecchio:2010mk,Dolag:2010ni}.  Despite the initial hindrance represented by the conformal conspiracy of conformally invariant Maxwell electromagnetism (EM) and conformally flat Friedmann-Lema\^itre-Robertson-Walker (FLRW) universes, abundant variations on this theme have been concocted by theorists, some of the simplest options revealing themselves as among the most elegantly successful in this competition.

So, why inflation?  In our opinion, besides the obvious ease with which such models can be studied -- in their most humble guise all one needs to do is to solve a homogeneous second order differential equation -- the reason is that the property which is perhaps the most intricate to frame within a consistent theoretical model is their pervasively wide coherence length, which can stretch well beyond the Mpc mark: cosmological inflation, unlike later time causal mechanisms, is automatically equipped with this feature.

Inflationary magnetogenesis is not without its list of drawbacks: many simple options suffer from a backreaction issue, see~\cite{Demozzi:2009fu,Kanno:2009ei,Urban:2011bu,Byrnes:2011aa}.  Very briefly, as inflation amplifies EM quantum fluctuations, the overall energy density of the latter ends up being comparable with that of the former, thereby implying the breakdown of perturbation theory.  When this occurs either inflation stops -- and if this happens too early then it does not generate enough entropy and curvature perturbations to explain today's Universe, or the magnetic fields cease to be amplified and remain too small today\footnote{Backreactions were first mentioned in an earlier work by Campanelli~\cite{Campanelli:2008kh}, where they were found to be unimportant thanks to a clever, albeit entirely arbitrary UV cutoff.}; the possibly simplest viable option~\cite{Martin:2007ue} appears to be successful only in the strong coupling (hence non perturbatively trustable) regime~\cite{Demozzi:2009fu,Kanno:2009ei}; furthermore, the counterproposal of~\cite{Caldwell:2011ra} also would not step down this cross~\cite{Barnaby:2012tk}.

In this battle for inflationary magnetogenesis there is, however, a survivor: this is the three-form driven inflation~\cite{Germani:2009iq,Koivisto:2009sd,Koivisto:2009ew}, which can be coupled in an elegant way to EM to amplify its initially tiny quantum fluctuations to sizeable amplitudes today, and at scales which easily exceed the Mpc scale~\cite{Koivisto:2011rm}.  In one line, the reason this particular model sails the winds where others float or sink, is that within it it is possible to carefully select which band of EM modes are to be amplified (for instance, the low momentum ones), thereby compressing all available energy density before backreaction into a handful of scales only.  This can be understood since the vector degrees of freedom in the three-form are excited by the coupling (consequently breaking down the duality with a scalar field inflation), resulting in new modes with a nontrivial dispersion relation. In particular, there appears a new mode that is exponentially growing over a certain range of wavemodes.

This is, however, not the end of the story.  Going beyond background dynamics, where only the EM field is treated as a small perturbation, a first order in the metric perturbations analysis was developed in~\cite{Bonvin:2011dt}.  Moreover, such coupling between the background scalar and EM can genuinely lead to non-Gaussianities in the Cosmic Microwave Background spectrum, see~\cite{Seshadri:2009sy,Barnaby:2011vw,Barnaby:2012tk,Brown:2010jd} and correlations between the scalar perturbations and the magnetic field~\cite{Caldwell:2011ra,Jain:2012ga}.  The EM field generated during inflation contributes in general an anisotropic term in the energy-momentum tensor, which marks the difference between the two scalar gravitational potentials (in longitudinal gauge), and has to be taken into account.  The conclusion is that such piece partly transmits to a constant curvature perturbation in the following radiation era: if this new term, whose absolute value depends on today's magnetic fields, becomes comparable to the background, it signals the breakdown of perturbation theory.

The scope of this work is to revisit this crucial issue for the particular case of the three form background.  We first review ``three-magnetogenesis'' in Sec.~\ref{threeEM}, and the appearance of the constant mode in Sec.~\ref{perturbed}; we apply this analysis to three-form inflation in~\ref{fatality}; with the tool-set at hand we briefly discuss non-Gaussianities generated in this scheme (Sec.~\ref{nong}); conclusions make up the closing Sec.~\ref{end}.

\section{Three-magnetogenesis}\label{threeEM}

We review and expand on the model first introduced in~\cite{Koivisto:2011rm}; in particular, we complete that work by presenting the derivation of the equations of motion in some detail.  We also discuss alternative prescriptions for the initial conditions of the system.

\subsection{Dynamics}

The dynamics and stability of three-form cosmology have been studied in~\cite{Koivisto:2009ew,Koivisto:2009fb,Boehmer:2011tp,DeFelice:2012jt} and considered also in the cases of nonminimal couplings to gravity~\cite{Germani:2009iq,Kobayashi:2009hj,Germani:2009gg} or to matter~\cite{Ngampitipan:2011se}.  In the following we will focus only on what is new here, that is the coupling to the electromagnetic field and the ensuing dynamics of the vector perturbations.  Trivially, the three-form field can be equivalently described by the dual vector field, and this sense
the models are examples of vector inflation~\cite{Golovnev:2008cf,Golovnev:2011yc,Watanabe:2009ct,Hervik:2011xm}.  The equivalent vector description would have a noncanonical form, but such vector models have been also studied recently~\cite{Dimopoulos:2009am,Dimopoulos:2009vu,Dimopoulos:2011pe}. 

Let us first specify the Lagrangian we study.  Let $A^\mu$ be the photon vector potential and $\B^{\mu\nu\rho}$ the three-form.  The canonical Lagrangian including both fields is\footnote{In terms of the dual vector, $V=V(6B^2)$ and $F^2(\B) \sim (\nabla\cdot B)^2$.}
\be\label{bare}
\lag_A+\lag_B = -\frac{1}{4}F^2(A)-\frac{1}{48}F^2(\B)-V(\B^2) \, ,
\ee
where the Faraday forms are computed from an $n$-form potential $N$ as $F(N)_{\mu_1\dots\mu_{n+1}}=(n+1)!\partial_{[\mu_1}N_{\mu_2\dots\mu_{n+1}]}$.  The components of the dual of the three-form are~\cite{Nakahara:2003nw}
\be
B_\alpha \equiv \frac{1}{6}\epsilon_{\alpha\beta\gamma\delta}\B^{\beta\gamma\delta}\,.
\ee
The most general Lorentz-invariant, quadratic, second order and $U(1)$ invariant coupling of the two fields 
is of the form
\be \label{inte}
\lag_{AB} &=& -\frac{1}{2}\lambda_1 F_{\mu\nu}(A)F^{\mu\nu}(B) \\
&&+\lambda_2(\nabla_\mu A^\mu)(\nabla_\nu B^\nu) + (\lambda_3+\lambda_4 R)A_\mu B^\mu + \lambda_5 A^\mu R_{\mu\nu} B^\nu + \lambda_6 A^\mu R_{\mu\nu\rho\sigma} \B^{\nu\rho\sigma} \, .\nn
\ee
Here $R_{\mu\nu\rho\sigma}$ is the Riemann tensor, and $R_{\mu\nu} \equiv R^\rho_{\;\mu\rho\nu}$ and $R \equiv R^\mu_{\;\mu}$ its Ricci contractions.  Only the first term respects the $U(1)$ symmetry of the photon. The second term breaks it only partially, leaving the reduced gauge invariance under the transformation $A \rightarrow A + \partial\phi$, where $\phi$ is a harmonic scalar field obeying $\Box\phi=0$. If $\lambda_3=0$, the remaining three terms vanish in flat space, and thus the standard predictions of QED are recovered regardless of the dynamics of the three-form.

We choose to work with only the $U(1)$ gauge invariant term, and set all $\lambda_j$ to zero, except for $\lambda_1$.  We expand the spatial part of the vector potential $A_\mu=(A_0,A_i)$ in terms of its transverse and longitudinal components as $A_i=A^T_i+\partial_i A^L$, where $\nabla\cdot {\bf A}^T = 0$.  The three-form is similarly decomposed $B_i=B^T_i+\partial_i B^L$.  Since the physical photons correspond to the transverse degrees of freedom, we are particularly interested in the vector perturbations.  The line element of the Friedmann universe including rotational perturbations can be parametrised as
\be
ds^2=a^2({\eta})[\dd\eta^2 + {\bf c}^T\cdot \dd{\bf x}\dd\eta - (\dd{\bf x}\cdot\nabla {\bf d}^T)\cdot \dd{\bf x}] \, . 
\ee
We will work in terms of ${\bf C}^T= - {\bf c}^T + {{\bf d}^T}'$, which includes the two gauge-invariant rotational degree of freedom of the metric perturbations.  The vector part of the Einstein-Hilbert Lagrangian coupled to the three-form is
\be\label{action0}
\lag_{EH+B}^{(v)} = \frac12 \left[ \frac{M_\text{P}^2 k^2}{2} {\bf C}^{T\,2} - \frac{V_{,X}}{X} \lp {\bf B}^T - X{\bf C}^T \rp^2 \right] \, ,
\ee
where we have used the background Friedmann equation and $X$ is the background value of the three-form field $X=\sqrt{B^2}$ in the notation of Ref.~\cite{Koivisto:2009fb}.

Let us now redefine $\lambda_1\equiv\lambda$. The relevant part of the action~(\ref{action0}) simplifies to
\be
S_A + S_{AB} &=& \frac12 \int \left\{ {\bf A}^T \lp - \partial^2_\eta + \Delta \rp \lp {\bf A}^T + 2\lambda {\bf B}^T \rp - A_0 \Delta \left[ A_0 - 2{A^L}' + 2\lambda \lp B_0 - {B^L}' \rp \right] \right. \nn\\
&& \left. - {A^L}' \Delta \left[ {A^L}' - 2\lambda \lp B'_0 - {B^L}' \rp \right] \right\}a^4(\eta) \dd^4 x \, .
\ee
Varying with respect to $A_0$, we get $A_0={A^L}'-\lambda(B_0-{B^L}')$, and the action further reduces to
\be\label{action}
\lag_{A+AB} = \frac12 \left[ {\bf A}^T \lp - \partial^2_\eta + \Delta \rp \lp {\bf A}^T + 2\lambda {\bf B}^T \rp + \lambda^2 \lp B_0 - {B^L}' \rp \Delta \lp B_0 - {B^L}' \rp \right] \, ,
\ee
where $\eta$ is conformal time, and $\Delta\equiv\delta^{ij}\partial_i\partial_j$.  The scalar polarisations of the photon are nondynamical, as expected.  Notice however that while the scalar part of the interaction decouples from the EM field as it should, it nontrivially contributes to the effective speed of sound of the three-form.

The equations of motion for the vector degrees of freedom become (omitting from now on the superscripts $T$, since all vectors are considered to be transverse)
\be
M_\text{P}^2 k^2 {\bf C} + 2V_{,X}({\bf B}-X{\bf C}) & = & 0 \, , \label{eom1} \\
V_{,X} ({\bf B}-X{\bf C}) - \lambda X(-\partial_\eta^2+\Delta){\bf A} & = & 0 \, , \label{eom2} \\
(-\partial_\eta^2+\Delta)({\bf A} + 2\lambda{\bf B}) & = & 0 \, . \label{eom_a}
\ee
We note that in the absence of the coupling, both the metric and three-form vector perturbations are nondynamical.  We can eliminate ${\bf C}$ and obtain a closed equation for the three-form perturbation in Fourier space:
\be\label{eom_b2}
\left[\partial_\eta^2 + \lp 1 + \frac{1}{f k^2} \rp k^2 \right] \cB = 0 \, ,
\ee
where
\be \label{f0}
f = 2\lambda^2 \frac{X}{V_{,X}} \lp \frac{2V_{,X}X}{M_\text{P}^2 k^2} - 1 \rp \, ,
\ee
and $\cB$ is the Fourier transform of $B$ and similarly we'll define $\cA$,
\be
{\cA}(\bk,\eta) \equiv \frac{1}{(2\pi)^{3/2}} \int \dd^3x A(\bx,\eta) e^{-i{\bf k}\cdot {\bf x}}\,, \quad
{\cB}(\bk,\eta) \equiv \frac{1}{(2\pi)^{3/2}}\int \dd^3x B(\bx,\eta) e^{-i{\bf k}\cdot {\bf x}}\,. 
\ee
Thus the three-form rotational modes propagate with a nontrivial dispersion relation; they can therefore in principle be significant even at large scales.

We are in the position to write an equation for the EM potential and the three-form only as $\cA'' + k^2 \cA = F(\cB)$, where $F(\cB) = -2\lambda(\partial_\eta^2+k^2) \cB$: the EM potential behaves as a harmonic oscillator driven by an external force.  Moreover, we can finally present the autonomous evolution equation for ${\bf A}$,
\be \label{eom_a2}
(-\partial_\eta^2+\Delta) \left[ 1 - f (-\partial_\eta^2+\Delta) \right] {\bf A} = 0 \, .
\ee
Each transverse degree of freedom in momentum space obeys
\ba \label{eom_a3}
\cA^{(4)} & + & 2\frac{f'}{f} \cA^{(3)} + \frac{1}{f} (f''-1+2k^2f) \cA'' + 2\frac{f'}{f} k^2 \cA' + k^2 (\frac{f''-1}{f} + k^2) \cA = 0 \, .
\ea

Fourth order time derivative appear in our equations, which may signal the presence of ghosts and/or negative norm states.  Since this is an effective field theory, and as such is only valid at low energies $k \ll V_0^{1/4}$ ($V_0$ being the UV mass scale of the model), as long as we confine ourselves to such energy range we are allowed to not worry about this potential drawback.

In order to be able to follow the analysis analytically, we consider de Sitter solutions with constant comoving field $X$. The two classes of fixed points~\cite{Koivisto:2009fb} are named $B$ (corresponding to $X^2=\pm\frac{2}{3}M_\text{P}^2$), and $C$ (corresponding to the minima of the potential).  Depending on the shape of the potential these can be stable or unstable; for this discussion what we need to know is that in both cases Eq.~(\ref{eom_a3}) much reduces to the simpler
\be\label{eomSim}
\cA^{(4)} - \lp\frac{1}{f_0} - 2k^2\rp \cA'' - k^2\lp \frac{1}{f_0} - k^2 \rp \cA = 0 \, ,
\ee
$f_0$ is given by Eq.~(\ref{f0}) evaluated at the fixed point.  We can readily solve this equation to obtain
\be
\cA(\eta) = \cA_1\cos{\lp k\eta \rp} + \cA_2\sin{\lp k\eta \rp} + \cA_3 e^{\Gamma k\eta} + \cA_4 e^{-\Gamma k\eta} \, ,
\ee
where $\cA_j$ are constant vectors and $\Gamma^2 + 1 \equiv 1/(f_0 k^2)$.  Hence, we have two additional solutions (as the equation is fourth order) which can be oscillatory if $\Gamma^2 < 0$, or, more interestingly, exponentially growing / decaying when $\Gamma^2 > 0$.

\subsection{Examples}

A perfect example of the features of this model is given by the exponential potential $V=V_0\exp{\lp - \xi X^2 / M_\text{P}^2 \rp}$.  At the fixed point $B$, which is stable as long as $\xi$ is positive, we have
\be\label{GammaDef}
\Gamma^2 = \frac{\kappa_\Lambda^2 - \kappa^2}{\Lambda^2 \kappa_\Lambda^2 + \kappa^2} \, ,
\ee
where we have defined $\Lambda^2 \equiv 8\lambda^2/3 / (1 - 8\lambda^2/3) \simeq 8\lambda^2/3$ and $k_\Lambda^2 \equiv 8\xi V / (3 M_\text{P}^2 \Lambda^2) \simeq \xi V / (\lambda^2 M_\text{P}^2)$.  Only for $k \leq k_\Lambda$ there exist an exponentially growing solution: $k_\Lambda$ acts in this case at the \emph{de facto} UV cutoff of the theory.  We have finally defined $\kappa \equiv k/\cH_e$ for convenience, where $\cH_e$ is the highest mode accessible by inflation -- we further denote $\cH \equiv a'/a$, where $a$ is the scale factor of the Universe and the prime stands for conformal time derivative.

The second fixed point $C$ can be an attractor if $\xi$ is negative.  In this case one sees that
\be
\Gamma^2 k^2 = \frac{\xi V_0}{\lambda^2 M_\text{P}^2} - k^2 \, .
\ee
Thus, when this point is an attractor, there is no instability of the vector modes. If we are inflating at a saddle point (i.e., $\xi>0$) vector modes will be unstable at scales
$k^2<\frac{\xi V_0}{\lambda^2 M_\text{P}^2}$.

The properties of the fixed points for many classes of potentials and their $\Gamma^2 +1$ are summarised in tables~\ref{tb2} and \ref{tb3}.

\begin{table*}
\begin{center}
\begin{tabular}{|c||c|c|}
\hline
$V(X)/V_0$ & $B$: stability & $B$: $\Gamma^2 + 1$   \\ \hline
$\exp(- \xi X/M_\text{P})$ & $B_{+}$ S for $\xi > 0$, $B_{-}$ S for $\xi < 0$ &$\frac{\xi V_0}{2\lambda^2 \lp \frac43 \xi V_0 \pm \sqrt{2/3} M_\text{P}^2 e^{\pm \xi \sqrt{2/3}} \rp}$  \\ \hline
$\exp(- \xi X^2 / M_\text{P}^2)$ & S for $\xi > 0$ & $\frac{\xi V_0}{\lambda^2 \lp \frac83 \xi V_0 + e^{-\frac23 \xi} M_\text{P}^2 k^2 \rp}$ \\ \hline
$(X/M_\text{P})^2$ & U & $\frac{V_0}{\lambda^2 \lp \frac83 V_0 - M_\text{P}^2 k^2 \rp}$ \\ \hline
$(X/M_\text{P})^{2n} , \, n>1$ & U &$\frac{n V_0}{\lambda^2 \lp \frac83 n V_0 - \lp \frac32 \rp^{n-1} M_\text{P}^2 k^2 \rp} $ \\ \hline
$\lp X^2 - C^2 \rp^2 / M_\text{P}^4$ & S for $C > \sqrt{2/3} M_\text{P}$ & $\frac{2 V_0 \lp \frac23 - C^2/M_\text{P}^2 \rp}{\lambda^2 \lp \frac{16}{3} V_0 \lp \frac23 - C^2/M_\text{P}^2 \rp - M_\text{P}^2 k^2 \rp}$ \\ \hline
\end{tabular}
\caption{\label{tb2} Stability of the de Sitter fixed points of type $B$ and the characteristic width of instability of the vector potential in four classes of models. U -- unstable; S -- stable.}
\end{center}
\end{table*}

\begin{table*}
\begin{center}
\begin{tabular}{|c||c|c|}
\hline
$V(X)/V_0$ &  $C$: stability & $C$: $\Gamma^2 + 1$  \\ \hline
$\exp(- \xi X/M_\text{P})$ 
& S & $\frac{\xi V_0}{2\lambda^2 X M_\text{P} k^2}$ \\ \hline
$\exp(- \xi X^2 / M_\text{P}^2)$ & S for $\xi < 0$ & $ \frac{\xi V_0}{\lambda^2 M_\text{P}^2 k^2}$ \\ \hline
$(X/M_\text{P})^2$ & S & $-\frac{V_0}{\lambda^2 M_\text{P}^2 k^2}$\\ \hline
$(X/M_\text{P})^{2n} , \, n>1$ & S & $0$\\ \hline
$\lp X^2 - C^2 \rp^2 / M_\text{P}^4$ & $C_1$ S, $C_2$ U & $\frac{2 V_0 C^2/M_\text{P}^2}{\lambda^2 M_\text{P}^2 k^2}$\\ \hline
\end{tabular}
\caption{\label{tb3} Stability of the de Sitter fixed points of type $C$ and the characteristic width of instability of the vector potential in four classes of models. U -- unstable; S -- stable.}
\end{center}
\end{table*}

One can therefore quite straightforwardly realise when the instability for the vector potential $\cA$, which is parametrised by $\Gamma^2$, is present, and where.  If large scales, $k\rar0$, are exponentially amplified, then $4\lambda^2 X^2/M_\text{P}^2 \leq 1$ or else we only have plane waves.  Also at large momenta we recover plane waves because $\Gamma^2\rar-1$.  Hence, there is in general a low energy band which is unstable given the above condition.

In addition to the turning point where $\Gamma^2$ changes sign, one should worry about one more singularity in $\Gamma^2$ which appears for the mode $k$ which nullifies its denominator.  The singularity could lie within the exponentially growing band, which is problematic since it would mean that to that mode corresponds infinite energy density.  One can preventively understand whether this will be the case by looking at the sign of $V_{,X}X$: if this is positive the divergence will appear; in the opposite case $V_{,X}X < 0$ the troublesome scale is beyond the UV cutoff.  This uncontrolled instability is consistently steered clear of in what follows.

We choose then to work with the exponential potential $V = V_0 \exp(-\xi X^2/M_\text{P}^2)$ because it does not require a very high degree of fine-tuning; a similar phenomenology arises in the Ginzburg-Landau case $V(X)=V_0(X^2-C^2)^2/M_\text{P}^4$, but only at the fixed point $B$.

\subsection{Some results}

The action~(\ref{action}) is canonical for the EM field; thus, since we regard the coupling to the three-form as a perturbation, we are allowed to proceed by quantising canonically~\cite{Barrow:2006ch}.  The solutions of interest are those which grow with conformal time, so we will choose $\cA_1 = \cA_2 = \cA_4 = 0$ -- this might appear contrived, but it is clear that, were these solutions kept, they would very rapidly loose any relevance since, in comparison to the growing $\cA_3$ piece, $\cA_1$ and $\cA_2$ are just plane waves, while $\cA_4$ decays exponentially.

At this point we need to fix the initial conditions (IC) of the dynamics for the EM vector potential $\cA$; we are presented with two possibilities: either we set them at some initial time (e.g.\ the beginning of inflation $\eta_i$), for all modes, or we normalise each mode to the Bunch-Davies vacuum $\sqrt{2k} \cA = e^{-ik\eta}$ at horizon crossing.  This last choice corresponds to a coupling which dynamically becomes relevant at some time, while being inefficient for smaller scales.  This option makes it clear that for all modes the specific IC do not matter, while they can be consistently chosen to be in vacuum at any early time without affecting the final result.  However, this demands a more elaborate theoretical motivation for the time-dependence in $\lambda$.

On the other hand, the former possibility is simpler, but perhaps not the most physical nor appealing one, for it introduces an explicit and very strong dependence on the IC of inflation, which is antithetic to what inflation is brought in for.  Moreover, there remains the formal question on how to consistently choose a vacuum state since the effective coupling is never turned off.  This, actually, can be fixed if we work in the small coupling regime: in this case it is consistent to choose the Bunch-Davies solution as the approximate IC.

One final consideration concerning the IC for the system.  We work with the simplifying assumption that $X$ is a constant, which implies that $f$ is so.  Then, the equation of motion for $\cB$~(\ref{eom_b2}) does not admit plane wave solutions at any time.  Nonetheless, it is quite easy to convince ourselves that in more realistic models, when $fk^2\rar\pm\infty$ as $k\eta\rar-\infty$ (which depends on the shape of the potential) then the rotational perturbation $\cB$ can, and in fact must, be initially in its Bunch-Davies vacuum, which in turn is realised for $\cA$.  That is, the exponent $\Gamma$ in reality is time-dependent as well, and flips from imaginary to real at a given time; we take the latter to be $\eta_i$ for simplicity.

{\bf IC at horizon exit} -- Let us work out the last case first.  The IC fix the value of $\cA_3$ and give us the following solution to work with:
\be
\cA &=& \frac{1}{\sqrt{2k}} e^{\Gamma (k\eta+1)} \quad \text{IC at } \eta = -1/k \,. ; \label{cAdef1}
\ee
Before moving on to the actual discussion, the key definitions we will need are the magnetic power spectrum
\be\label{deltaB}
\delta_B^2 \equiv \frac{k^5 |\cA|^2}{4\pi^2 a^4} \, ,
\ee
where $a$ is the scale factor of the Universe, and the energy density in electromagnetic field:
\be\label{energy}
\rho_\eb = \frac{1}{4\pi^2 a^4} \int \frac{\dd k}{k} k^3 \left[ |\cA'|^2 + k^2 |\cA|^2 \right] \, ,
\ee
where the first term is related to the electric field, and the second one is the magnetic energy density proper.  Recall that for a de Sitter solution conformal time spans the range $[-\infty,0]$, and $\cH = -1/\eta$.  We can immediately write down the corresponding expressions for the energy density associated with the EM field at the end of inflation as
\be
\rho_\eb &=& \frac{\cH_e^4}{8\pi^2 a_e^4} \int\dd\kappa \kappa^3 \lp \Gamma^2 + 1 \rp e^{2\Gamma(1-\kappa)} \, . \label{energy1}
\ee
where of course $\kappa_e = 1$ by definition, and where we can clearly ignore the second piece in the exponent, being greatly smaller than the first, positive, contribution.  Also, we work with small $\lambda$, which means $\Lambda\ll1$ and $\Gamma\gg1$; this implies that in general $E^2 \gg B^2$.

The energy density Eq.~(\ref{energy1}) is expressible in closed form.  The limits of integration are $\kappa = \kappa_i = \exp(-N)$, where $N$ is the total number of e-folds inflation lasts for, and $\kappa = \kappa_\Lambda$, beyond which the solution~(\ref{cAdef1}) does not apply any more.  The final expression 
to be evaluated at those limits, reads
\be\label{energy1solved}
\rho_\eb = \frac{\cH_e^4}{8\pi^2a_e^4} \frac{\kappa_\Lambda^4}{2} \!\!&&\!\! \left\{ \frac{e^{2\Gamma}}{1+\Gamma^2} \lp 1+\Lambda^2 + \Lambda^4 \rp + i e^{2i} \left[ 1 + \Lambda^2 + (1-i) \Lambda^4 \right] E_i \lp 2\Gamma - 2i \rp \right. \nn\\
&& \;\left. - i e^{-2i} \left[ 1 + \Lambda^2 + (1+i) \Lambda^4 \right] E_i \lp 2\Gamma + 2i \rp \right\} \, .
\ee
This expression is not particularly illuminating in this form, but can be expanded for small $\Lambda$ to yield
\be\label{energy1final}
\rho_\eb &=& \frac{\cH_e^4}{8\pi^2a_e^4} \frac{\kappa_\Lambda^4}{2} \Lambda^3 e^{2/\Lambda} = \frac{k_\Lambda^4}{16\pi^2 a_e^4} \Lambda^3 e^{2/\Lambda} \, .
\ee
Notice how the cutoff scale $k_\Lambda$ appears, where normally one has $\cH_e$, and how the exponent is inversely proportional to $\Lambda$.  In the same fashion, we can express the spectrum as seen today, diluted by a factor $(a_e/a_0)^4$, as
\be\label{deltaB1final}
\delta_B^{2,0} = \frac{k^4}{8\pi^2} e^{2\Gamma} \simeq \frac{k^4}{8\pi^2} e^{2/\sqrt{\Lambda^2 + (k/k_\Lambda)^2}} \, .
\ee
In order to have a low-scale cutoff the scale of the three-form potential needs to be low.  Moreover, since $\Gamma \propto 1/\lambda$ we want a small coupling constant for the most efficient amplification.

\begin{figure}[ht]
\centering
\includegraphics[width=0.45\textwidth]{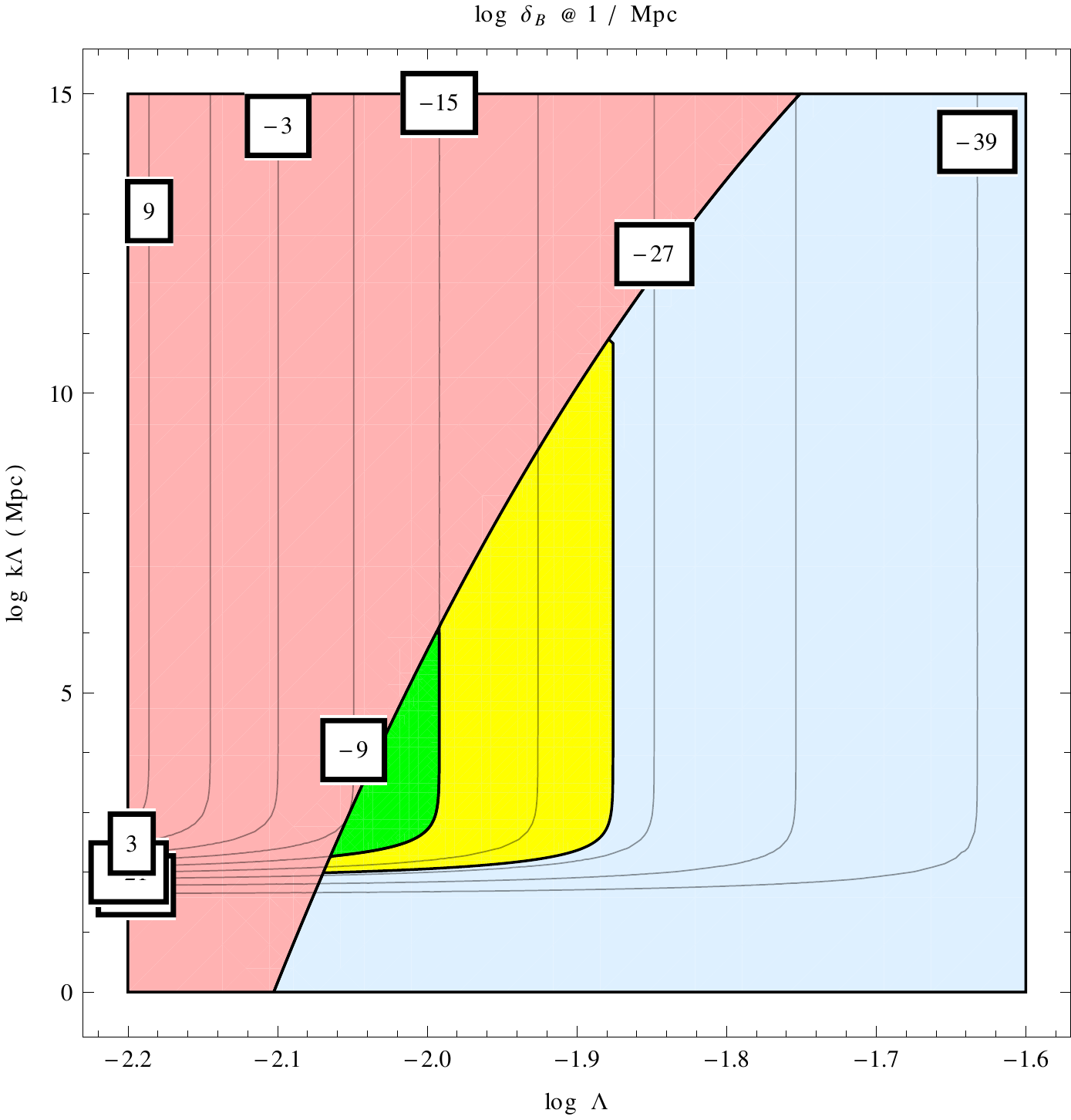}
\includegraphics[width=0.45\textwidth]{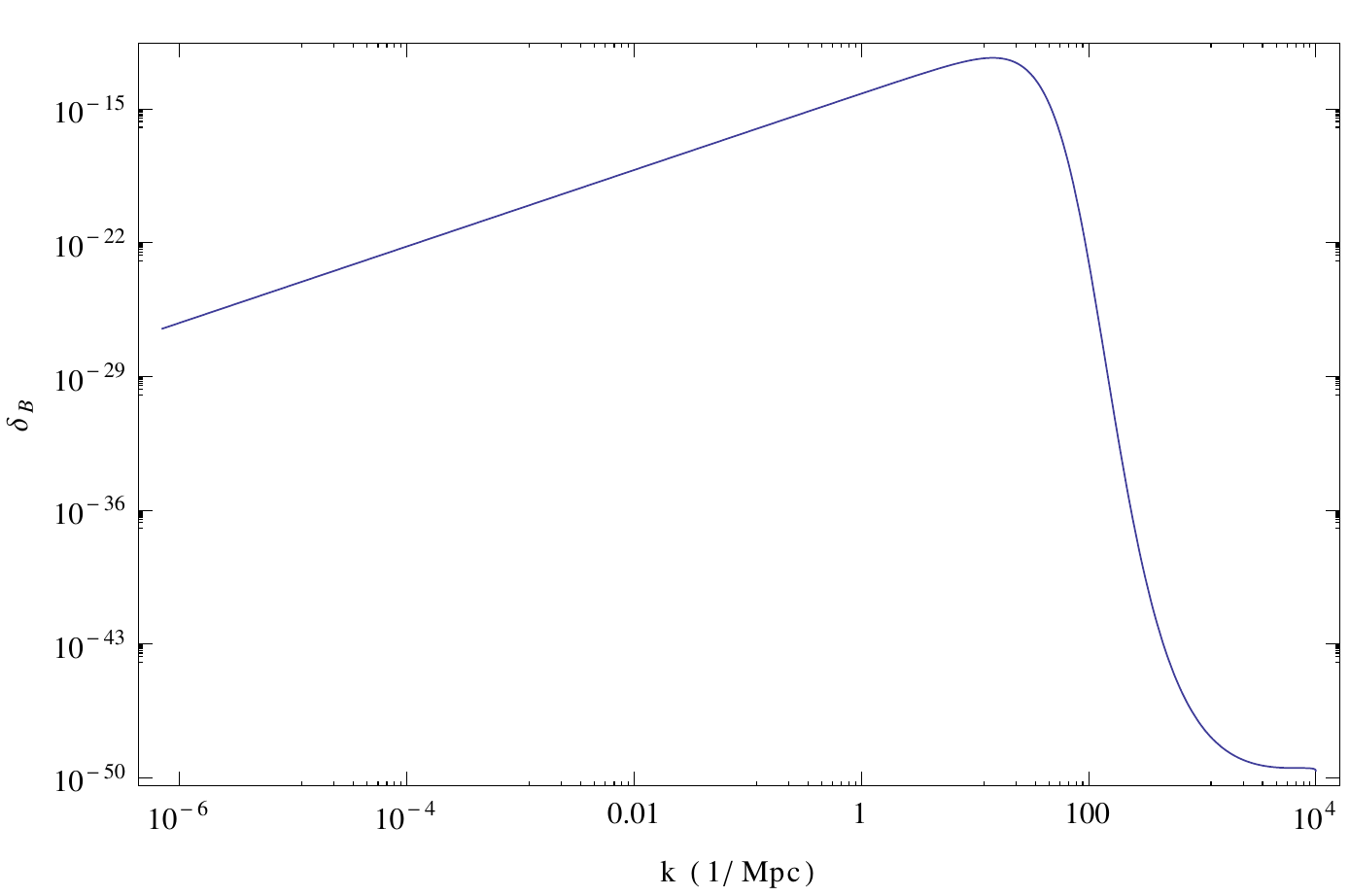}
\caption{\emph{Left}.  Allowed (in light blue) and disallowed (in light red) regions in the $(\log\Lambda, \log\lp k_\Lambda \Mpc \rp)$ parameters space; in the green region the generated magnetic field can explain magnetisation at and beyond the Mpc scale, while in the yellow region some degree of dynamo would be required.  Superimposed are contours of constant $\log\delta_B^0$ today, at 1/Mpc (in Gauss units).  \emph{Right}.  The power spectrum today, $\delta_B^0$ (Gauss), for $\Lambda=10^{-2}$ and $k_\Lambda = 10^4 / \Mpc$.}
\label{fig1}
\end{figure}

In Fig.~\ref{fig1}, left panel, we present the regions in the parameter space $(\Lambda, k_\Lambda)$ for which the ratio $\rho_\eb / \rho_X$ stays below one at the end of inflation (blue region) and goes beyond one, thereby leading to instability in the de Sitter background (red region).  Superimposed are the contours for a given value of the magnetic power at 1/Mpc today, where the regions for which $\delta_B^0 \geq 10^{-15}$ Gauss and $\delta_B^0 \geq 10^{-25}$ Gauss are in evidence (green and yellow, respectively).  These are values that correspond to the magnetic fields observed in the intergalactic medium, and a typical value for a successful seed to be fed the magnetohydrodynamical plasma at late times.  The right panel of Fig.~\ref{fig1} plots the power spectrum today; notice again the knee at high energy and the following very rapid decay.  Notice how such a spectrum also nicely helps in avoiding the stringent constraints which late time cosmology poses on smaller scale fields~\cite{Caprini:2011cw}.  These figures were obtained for around $N=66$ e-foldings of inflation, $\cH_e/a_e \approx 10^{13}$ GeV, and assuming the Universe was always dominated by radiation from the end of inflation onwards, but they are by construction fairly independent on these specifics.

\begin{figure}[ht]
\centering
\includegraphics[width=0.45\textwidth]{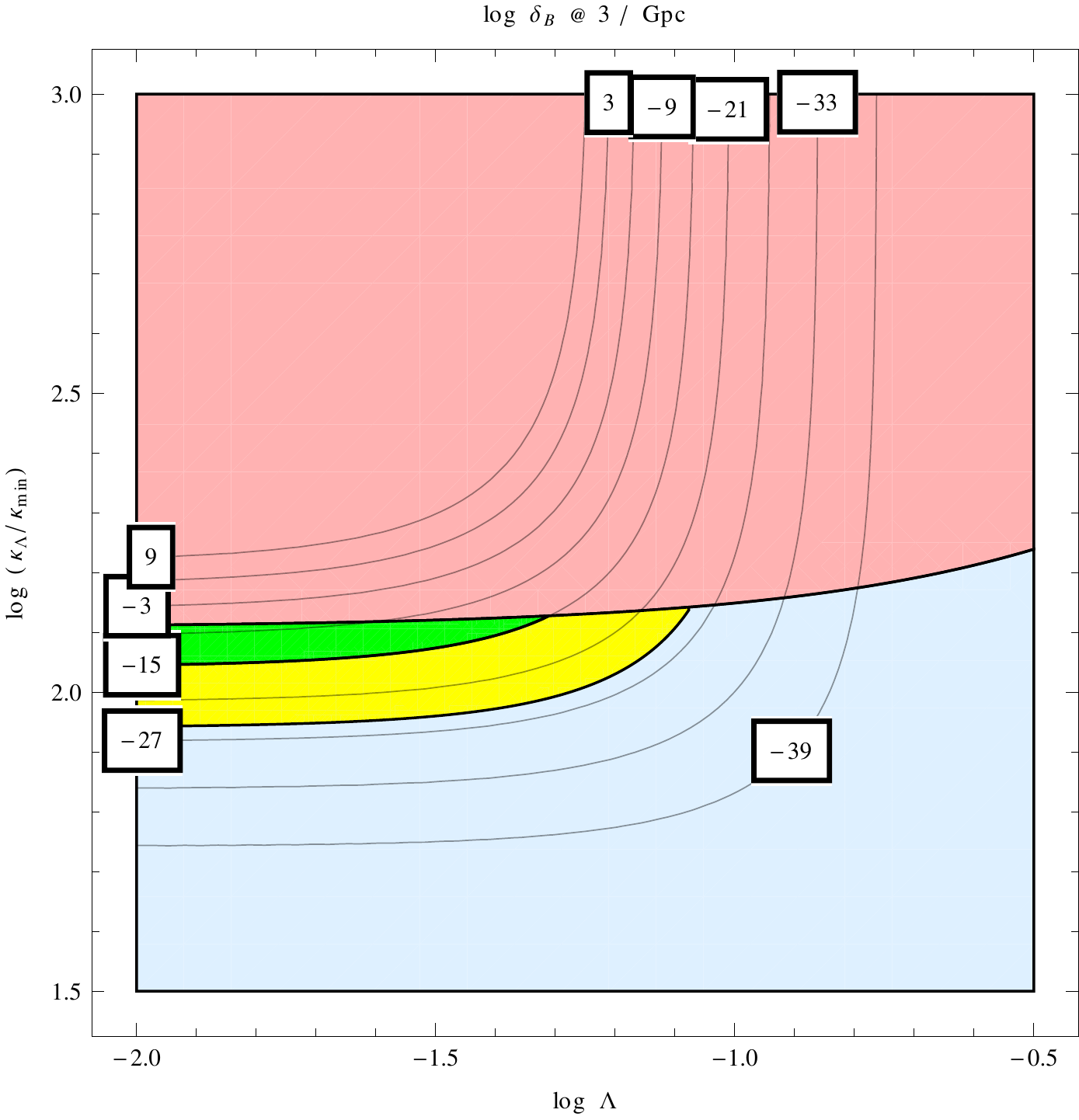}
\includegraphics[width=0.45\textwidth]{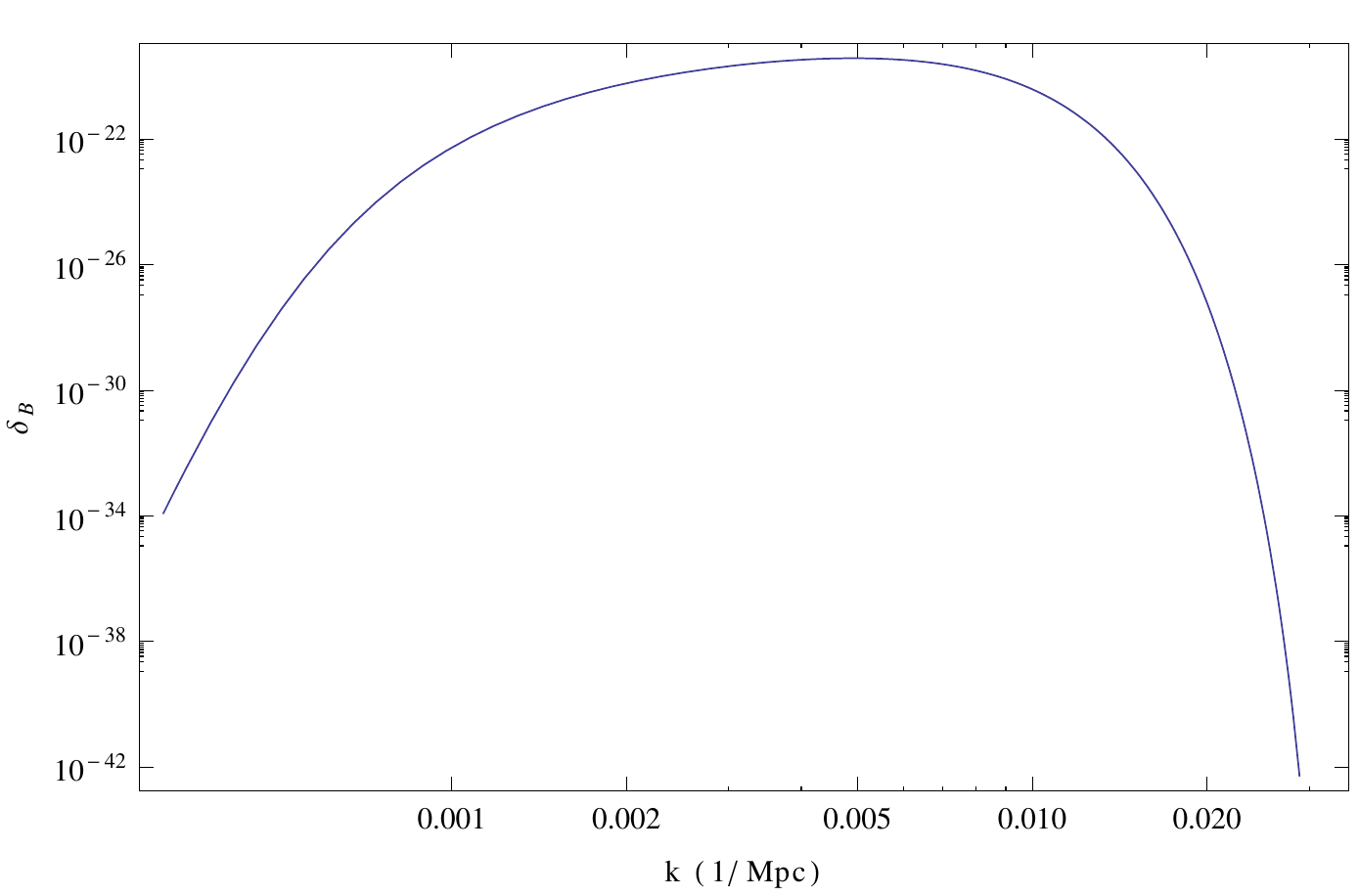}
\caption{\emph{Left}.  Allowed (in light blue) and disallowed (in light red) regions in the $(\log\Lambda, \log\lp k_\Lambda/k_\text{min} \rp)$ parameters space; in the green region the generated magnetic field can explain magnetisation at and beyond the Mpc scale, while in the yellow region some degree of dynamo would be required.  Superimposed are contours of constant $\log\delta_B^0$ today, at 3/Gpc (in Gauss units).  \emph{Right}.  The power spectrum today, $\delta_B^0$ (Gauss), for $\Lambda=10^{-2}$ and $k_\Lambda / k_\text{min} = 10^2$.}
\label{fig2}
\end{figure}

{\bf IC at the beginning of inflation} -- For clarity, we discuss separately the calculation with the alternative prescription for the initial conditions.  Let us set: 
\be
\cA &=& \frac{1}{\sqrt{2k}} e^{\Gamma k(\eta-\eta_i)} \quad \text{IC at } \eta = \eta_i \, . \label{cAdef2}
\ee
The energy density is then computed from the integral
\be
\rho_\eb &=& \frac{\cH_e^4}{8\pi^2 a_e^4} \int\dd\kappa \kappa^3 \lp \Gamma^2 + 1 \rp e^{2\Gamma\kappa(1/\kappa_i-1)} \, . \label{energy2}
\ee
The integral~(\ref{energy2}) instead is not simply dealt with, and we must resort to numeric methods to solve it.  In Fig.~\ref{fig2}, left, we present the same regions in the parameter space $(\Lambda, k_\Lambda)$ for $\exp (- \xi X^2)$.  Fig.~\ref{fig2}, right, plots the power spectrum today, for $\Lambda=10^{-2}$ and $k_\Lambda / k_\text{min} = 10^2$.

As one can see by looking at these figures, and as it can be expected based on Eq.~(\ref{cAdef2}), there is some fine tuning in this case; there is a very strong sensitivity (doubly exponentially in terms of the total number of e-folds $N$) to the initial time, and only scales not too far from the first ones available to inflation can be amplified efficiently without leading to a catastrophic backreaction effect.  We picked the representative scale $k$ = 3/Gpc, and the largest mode is only around $1/\cH_0 \approx$ 4Gpc.  By looking at the spectrum we realise how only a handful (not more than a couple, in this case) of orders of magnitude in $k$ will be amplified efficiently; for instance, the strength at 10/Gpc is $10^{-21}$ Gauss, but drops to $10^{-47}$ Gauss at just 30/Gpc.  The scales we amplify therefore heavily rely on what the largest scale is.  With the alternative prescription of a time-dependent coupling the sensitivity to initial conditions can in principle be eliminated.  Notice that here inflation lasts about 60 Hubble times.

From the allowed regions in Fig.~\ref{fig2} it seems that we could push the coupling $\lambda\simeq\Lambda$ to arbitrarily small values, and still be able to obtain sizeable power at the scales we need.  This is only true in part: since $\lambda$ also enters the UV cutoff, and as $\lambda\rar0$ one needs to keep $\lambda^2 k_\Lambda^2 \simeq \xi V_0 e^{-2\xi/3} / M_\text{P}^2$ fixed, the window of modes which can be boosted has to be kept narrow.  Lastly, one could consider models with either very high or very low potential slope parameters $\xi$: this would help reconciling suitable UV cutoffs with weaker interaction strengths $\lambda$.

Having now shown that the three-form magnetogenesis can avoid the backreaction problem at the zeroth order, we turn to consider the linear perturbations more carefully.

\section{The framework for perturbations}\label{perturbed}

We now consider the scalar sector of metric perturbations, and review the results of~\cite{Bonvin:2011dt} following closely their presentation.

\subsection{Metric perturbations and anisotropies}

The starting point of the analysis of~\cite{Bonvin:2011dt} is the perturbed metric in longitudinal gauge:
\be
ds^2 = a^2 \left[ (1+2\Phi) \dd\eta^2 - (1-2\Psi) \dd\bx^2 \right] \, .
\ee
In order to obtain the perturbed Einstein field equations $\delta G_{\mu\nu} = 8\pi G \delta T_{\mu\nu}$ we need an expression for the energy-momentum tensor (the Planck mass is $8\pi G \equiv M_{\rm P}^2$); we can decompose it with respect to the velocity of the FLRW background, $u^\mu =a^{-1}(1-\Phi,{\bf 0})$, and to the metric $g_{\mu\nu}$ as is customary as:
\be
T^{\mu\nu}_{\eb}=(\rho_\eb +p_\eb) u^\mu u^\nu - p_\eb g^{\mu\nu} + u^{\mu}q^{\nu}_{\eb} + u^{\nu}q^{\mu}_{\eb} +\Pi^{\mu\nu}_{\eb} \, , \label{Tem}
\ee
where we take the EM field to be half order in perturbation theory, where, thence, only the unperturbed parts of the velocity $u^\mu$ and metric $g_{\mu\nu}$ will appear.

In standard inflation the background fluid is described by a scalar field, $\cph$, expanded as $\cph(\eta,\bx) = \hat\cph(\eta) + \delta\cph(\eta,\bx)$, with a standard kinetic term and potential.  In this case, the background equations of our system are
\be
&& 4\pi G \hat\cph^{'2} = \cH^2 - \cH' \, ,\label{back}\\
&& \hat\cph'' + 2\cH \hat\cph' + a^2 V_{,\cph} = 0 \, .\label{backeom}
\ee

It is straightforward to write down the perturbed Einstein equations now, which, in Fourier space and in the presence of an electromagnetic field are:
\be
&& 3\cH \Psi' + (2\cH^2 + \cH') \Phi + k^2 \Psi = -4\pi G\Big(\hat\cph' \delta\cph' + V_{,\varphi} a^2\delta\cph\Big) - 4\pi Ga^2 \rho_\eb \, , \label{psiprime}\\
&& \Psi''+ 2\cH \Psi' + \cH \Phi' + (2 \cH^2 + \cH') \Phi - \frac{k^2}{3} (\Phi-\Psi) = 4\pi G \Big( \hat\cph' \delta\cph' - V_{,\cph} a^2 \delta\cph\Big) + 4\pi G a^2 p_\eb \, , \nn\\\label{psidprime}\\
&& \Psi' + \cH \Phi = 4\pi G \hat\cph' \delta\cph - 4\pi G a \, {\rm i} \, \frac{k^j}{k^2}\, q_{\eb\, j} \, , \label{psiq}\\
&& k^2 (\Phi-\Psi) = -8\pi G a^2 \pisubs \, . \label{phiminuspsi}
\ee
We define $\rho_\eb(\bk)$, $p_\eb(\bk)$ and $q_{\eb\, j}(\bk)$ as the EM field energy density, pressure and Poynting vector (all these in momentum space).  We also define $\pisubs(\bk)\equiv -3 \hat k^i\hat k_j \Pi^{\ j}_{\eb\, i} (\bk)/2$ as in~\cite{Bonvin:2011dt}.  Notice that the background equation~(\ref{back}) has been used here.  The anisotropic contribution $\pisubs$ in Eq.~(\ref{phiminuspsi}) is what prevents the usual identification $\Phi = \Psi$.

Eliminating $V_{,\cph}$ through the background equation~(\ref{backeom}), using Eq.~(\ref{psiq}) to obtain $\hat\cph'\delta\cph$, and finally adding~(\ref{psiprime}) and~(\ref{psidprime}) we derive an equation (Bardeen) for $\Psi$ alone as
\be
\label{psi}
\Psi'' + 2\left( \cH - \frac{\hat\cph''}{\hat\cph'} \right) \Psi' + \left( 2\cH' - \frac{2\cH\hat\cph''}{\hat\cph'} + k^2 \right) \Psi = \source \, ,
\ee
where every $\Phi$ instance has been replaced by $\Psi$ through Eq.~(\ref{phiminuspsi}).  The source term $\source$ collects all contributions from the EM field, and reads
\be
\label{source}
\source &=& 8\pi G a^2 \left\{ \cH \frac{(a^2 \pisubs)'}{a^2 k^2} + \left[ 2\left( \cH' - \cH \frac{\hat\cph''}{\hat\cph'} \right) - \frac{k^2}{3} \right] \frac{\pisubs}{k^2} \right. \nn\\
&& - \left. \frac{1}{2}(\rho_\eb - p_\eb) + \left( 2\cH + \frac{\hat\cph''}{\hat\cph'} \right) \frac{ {\rm i}\, k^jq_{\eb\, j}}{k^2 a} \right\} \, .\nn\\
\ee

Since we are interested in magnetic fields which today have extremely large wavelengths, we focus our attention to scales for which $x \equiv k|\eta|\ll1$ (the conformal time in almost de Sitter background being negative) and we develop everything in the slow roll approximation; the slow roll parameters themselves are defined as
\be
\cH^2 - \cH' = \ep_1 \cH^2 \quad \mbox{and} \quad \ep_1' = 2\ep_1 (2\ep_1 + 3\ep_2) \cH \, .\label{slro}
\ee
These can be recast in another useful form for our subsequent computations as
\be
\cH' = \cH^2 (1-\ep_1) \quad \mbox{and} \quad \cH'' &=& 2 \cH^3 (1-2\ep_1-\ep_1^2-3\ep_1\ep_2) \, .\label{slroh}
\ee
Time, scale factor, and the Hubble parameter are thence related as $a \propto 1/\eta^{1+\ep_1}$ and $\cH \simeq -(1+\ep_1)/\eta$.  

Putting all this together, we arrive at the master equation
\be\label{psiroll}
\frac{\dd^2 \Psi}{\dd x^2} + \frac{2(\ep_1 + 3\ep_2)}{x} \frac{\dd \Psi}{\dd x} + \left[1 - \frac{2(2\ep_1 + 3\ep_2)}{x^2} \right] \Psi = \cS_\eb \, ,
\ee
where
\be\label{sourceLS}
\cS_\eb \simeq \frac{8\pi G a^2 \cH^2 }{k^4} \left[ \frac{\pisubs'}{\cH} + 2\left( 1 - 2\ep_1 - 3\ep_2 \right) \pisubs - \frac{k^2}{3} \pisubs - \frac{k^2}{2} (\rho_\eb - P_\eb) + 3 k^2 Q_\eb \right] \, ,
\ee
where we defined
\be\label{Qdef}
Q_\eb \equiv \frac{ {\rm i}\, \cH k^jq_{\eb\, j}}{k^2 a} \, ,
\ee
and $\cS_\eb \equiv \source/k^2$.  We have expanded everything to order $x^2$ inside the parentheses, but have kept the the first nonzero slow roll order only for the term $\pisubs/x^2$, the reason for which will become clear below -- this is a consistent expansion for our purposes.

It is possible to show~\cite{Bonvin:2011dt} that the flow $Q$ is related to the other quantities in the source by
\be\label{Qcons}
\frac{Q_\eb'}{\cH} = \frac{2}{3} \pisubs - P_\eb - \frac{4\cH^2-\cH'}{\cH^2} Q_\eb \, ;
\ee
Later we will also use the fact that, of course, $P_\eb = \rho_\eb/3$.

\subsection{The source unveiled}

In order to solve Eq.~(\ref{sourceLS}) one needs to compute $\pisubs$; this can be done if the time evolution of the source generating the anisotropy is known, which in turn, in the EM case, can be read off the solutions of the equation of motion for the EM vector potential coupled to the inflaton.  The simplest, and analytically solvable, model in this sense is given by 
\be\label{actionem}
\cL = - \frac{1}{4} f^2(\cph) F^{\mu \nu}F_{\mu \nu} \, ,
\ee
In Coulomb gauge $A_0=0$, $\partial_j A^j=0$ it is straightforward to solve the Euler-Lagrange equations for the Fourier modes\footnote{Notice that in order to apply our formul\ae\ in the previous and following sections we need to canonically normalise the field.} $f\cA$
\be
\left(f\cA\right)'' + \left( k^2 - \frac{f''}{f} \right) f\cA = 0 \, , \label{eqA}
\ee
where the coupling function $f(\cph)$ is parametrised as a power law as:
\be\label{f}
f(\eta)\propto\eta^\gamma \, .
\ee 

Typically the value of $\gamma$ is restricted to be within $[-2,2]$, both because most quantities (power spectra, energy densities, etc.) are well-defined in this range, and the EM field remains subdominant during sufficiently long-lasting
inflation.  In fact, as for the latter, it can be shown that slightly larger values of $|\gamma|\lesssim2.2$ are allowed (see for instance~\cite{Demozzi:2009fu}).

Skipping the details of quantisation, the solutions to~(\ref{eqA}) are given by
\be
\cA(k,\eta) = \sqrt{\frac{x}{k}} \left[ C_1(\gamma) J_{\gamma - 1/2}(x) + C_2(\gamma) J_{-\gamma + 1/2}(x) \right] \, ,\label{Asolution}
\ee
where $J_\nu$ are Bessel function of order $\nu$,
\be\label{bigC}
C_1(\gamma) = \sqrt{\frac{\pi}{4}} \frac{e^{- {\rm i} \pi\gamma/2}}{\cos\pi\gamma} \, , \text{ and} \quad C_2(\gamma) = \sqrt{\frac{\pi}{4}} \frac{e^{ {\rm i} \pi(\gamma+1)/2}}{\cos\pi\gamma} \, .
\ee
The long wave expansion for this solution reads
\be\label{Aapp}
\cA(k,\eta) \simeq \frac{1}{\sqrt{k}} \Big[c_1(\gamma) x^\gamma + d_1(\gamma) x^{\gamma + 2} + c_2(\gamma) x^{1 - \gamma} + d_2(\gamma) x^{3 - \gamma} \Big] \, ,
\ee
with
\be
&& c_1(\gamma) = \frac{{\rm e}^{- {\rm i} \pi \gamma/2}}{\cos(\pi\gamma)} \frac{\sqrt{\pi/4}}{2^{\gamma - \frac{1}{2}} \Gamma(\gamma + 1/2)} \, , \quad d_1(\gamma) = \frac{c_1(\gamma)}{\gamma + 1/2} \, , \\
&& c_2(\gamma) = \frac{{\rm e}^{{\rm i} \pi (\gamma + 1)/2}}{\cos(\pi\gamma)} \frac{\sqrt{\pi/4}}{2^{\frac{1}{2} - \gamma} \Gamma(3/2 - \gamma)} \, , \quad d_2(\gamma) = \frac{c_2(\gamma)}{3/2 - \gamma} \,.
\ee

Every term of the source $\cS_\eb$ can be extracted from the general expression
\be
\label{Tmunu}
T^{\quad\nu}_{\eb\;\mu} = f^2 \left(F_{\mu\lambda} F^{\nu\lambda} - \frac{1}{4}\delta_{\mu}^{\ \nu} F^2 \right) \, ,
\ee

At this point, overflying the details of the hard part of the calculation, three steps are necessary in order to derive a useful expression for the source term.  First, since it can be shown that only the time dependence of this term is relevant for us, the most direct way to do so is to compute the power spectrum
\be\label{timeevol}
\langle0| \pisubs^\dagger(\bq,\eta) \pisubs(\bk,\eta)|0\rangle = (2\pi)^3 P_\Pi(k,\eta) \delta(\bq-\bk) \, ,
\ee
and then identify the classical source with the square root of this; quoting again~~\cite{Bonvin:2011dt}, we have
\be\label{ppidef}
P_\Pi(k,\eta) = \frac{f^4}{|\eta|^3} \int_0^{1} y^2 \dd y \, \sum_i \si_i(\gamma) P_i(y,\eta) P_i(|x-y|,\eta) \, , \quad i\equiv \{E, B, EB \} \, ,
\ee
where the order one $\si_i$ coefficients are mildly $\gamma$ dependent (within the region $|\gamma|\lesssim2$ only), and we have ignored a bunch of factors $2\pi$ for simplicity.  The power spectra appearing in this convolution are as usual defined as
\be
P_B &=& \frac{k^2}{f^2 a^4} |\cA|^2 \, , \label{PB}\\
P_E &=& \frac{1}{a^4} \left| \left( \frac{\cA}{f} \right)' \right|^2 \, ,\label{PE}\\
P_{EB} &=& \frac{k}{f a^4} \left( \frac{\cA}{f} \right)' \cA^* \, .\label{PEB}
\ee

All that matters in this expression is the final time dependence, which can be calculated using some rough approximation for the convolution integral for all values of $\gamma \in [-2,2]$ (see the details in the original papers~\cite{Bonvin:2011dt}) and cast in a simple and illuminating form
\be\label{ppi0}
P_\Pi(k,\eta) \simeq \hat P(k) |\eta|^{2\alpha} \, ,
\ee
with
\be\label{whatsalpha}
\alpha(\gamma) = \min\{4-2|\gamma|,3/2\} \, .
\ee
This means that
\be\label{eqpis}
\pisubsp = -\alpha \, \cH \, \pisubs \, ,
\ee
whence
\be\label{whatspip}
\cS_\eb \simeq \frac{3}{x^4} \left[ (2-\alpha+4\ep_1-3\alpha\ep_1-6\ep_2)\pisubs - \frac{x^2}{3} \pisubs - \frac{x^2}{2} (\rho_\eb - P_\eb) + 3 x^2 Q_\eb \right]
\ee
($\rho_\cph$ is the energy density of $\cph$).

\subsection{Solving for the Bardeen potentials}

It is now possible to solve the Bardeen equation~(\ref{psiroll}); the solutions are obtained as a linear combination of two homogeneous solutions, which are $\Psi_1(x)=x^pJ_\nu(x)$ and $\Psi_2(x)=x^pJ_{-\nu}(x)$, with $p=1/2+\mathcal{O}(\ep_1,\ep_2)$ and
$\nu=1/2+\mathcal{O}(\ep_1,\ep_2)$, and a particular solution of the inhomogeneous equation, obtained through the two homogeneous solutions and the Wronskian
\be\label{wronskian}
\cW(x) = \Psi_1(x) \Psi_{2,x}(x) - \Psi_{1,x}(x) \Psi_2(x) = - \frac{2}{\pi} \sin(\nu\pi) x^{2p-1} \, ,
\ee
as
\be\label{psiinhW}
\Psi_{\rm inh}(\bk, x) = w_1(\bk) \Psi_1(x) + w_2(\bk) \Psi_2(x) \, ,
\ee
where
\be\label{w1w2}
w_1(\bk) \equiv - \int\dd z \frac{\Psi_2(z) \cS_\eb(\bk, z)}{\cW(z)} \quad , \quad w_2(\bk) \equiv \int\dd z \frac{\Psi_1(z) \cS_\eb(\bk, z)}{\cW(z)} \, .
\ee
Thence:
\be
\label{psiinh}
\Psi_{\rm inh}(\bk, x) = & \frac{\pi}{2\sin(\nu\pi)} \int_{x_{\rm h}}^x \dd \ln z \left(\frac{x}{z}\right)^p \!&\! \left[J_\nu(x) J_{-\nu}(z) - J_\nu(z) J_{-\nu}(x) \right] z^2 \cS_\eb(\bk, z) \nn\\
\simeq & \frac{1}{2\nu} \int_{x_{\rm h}}^x \dd \ln z \left(\frac{x}{z}\right)^p \!\!\!\!&\!\!\!\! \left\{\left(\frac{x}{z}\right)^\nu \left[ 1+\frac{z^2}{4\nu-4}-\frac{x^2}{4\nu+4} \right] \right. \\
&&\!\!\!\! - \left. \left(\frac{z}{x}\right)^\nu \left[ 1-\frac{z^2}{4\nu+4}+\frac{x^2}{4\nu-4} \right] \right\} z^2 \cS_\eb(\bk, z) \, , \nn
\ee
plus higher order terms; we keep the second order terms in the expansion since, as we will see shortly, the first term does not contribute to the final potentials.  Notice here that, unlike the convolution Eq.~(\ref{ppidef}), where for convenience we rewrote the momentum $k$ in terms of the dimensionless $x$, but the integral was a momentum integral, here we are integrating over conformal time $\eta$ -- this is also why we chose $y$ as dummy variable there, and $z$ here.  This time integral runs from horizon exit $x_{\rm h} \simeq 1$ as this is when the mechanism becomes operational; we also have expanded the integrand for large scales $x\ll1$.

Focussing only on the ``strictly'' inhomogeneous solution at $z\rar x$, and ignoring the homogeneous solutions as they are subdominant, one can write then
\be
\Psi_{-} & \simeq & \hat b(k) - \frac{3}{3-\alpha} \left[1+\frac{5-3\alpha}{3-\alpha}\ep_1 \right] \frac{\pisubs}{x^2 \rho_\cph} - \frac{1}{(3-\alpha)(1-\alpha)} \left[ \frac{\pisubs}{\rho_\cph} - \frac{6-\alpha}{\alpha} \frac{\rho_\eb}{\rho_\cph} \right] \, , \label{psiminusO}\\
\Phi_{-} & \simeq & \Psi_{-} - 3(1+2\ep_1)\frac{\pisubs}{x^2 \rho_\cph} \, ; \label{phiminusO}
\ee
$\hat b(k)$ denotes the contribution from the scalar field fluctuations at large scales.  The last step is the quantum-to-classical transition, in which the quantum operators of~(\ref{psiminusO}) and~(\ref{phiminusO}) are transformed into the squares of their power spectra, leading directly to
\be
\Psi_{-}(k) & \simeq & b(k) - \frac{3}{3-\alpha} \left[1+\frac{5-3\alpha}{3-\alpha}\ep_1 \right] \frac{\ompi}{x^2} k^{-3/2} \nn\\
&& - \frac{1}{(3-\alpha)(1-\alpha)} \left( \ompi k^{-3/2} - \frac{6-\alpha}{\alpha} \omrho k^{-3/2} \right) \, , \label{psiminus}\\
\Phi_{-}(k) & \simeq & \Psi_{-} - 3(1+2\ep_1) \, \frac{\ompi}{x^2} k^{-3/2} \, , \label{phiminus}
\ee
where the dimensionless energy density fraction $\ompi$ is conveniently defined as
\be\label{ompdef}
\left[\ompi(k,\eta)\right]^2 \equiv \frac{k^3 P_\Pi}{\rho^2_\cph} \, ,
\ee
and similarly for $\omrho$.

Once inflation ends, the Universe enters the reheating phase, which is (hopefully) followed by epochs of matter or radiation domination, once the decay products of the inflaton (hopefully) thermalise.  The reheating epoch, although now yet completely understood, typically lasts a short (on Hubble scales) time, which means that for very large-scale modes which are sitting outside the Hubble radius, this process is essentially instantaneous and not influential on their dynamics.  Hence, in order to tailgate the destiny of the Bardeen potentials to our times we need to match the inflationary solutions~(\ref{psiminus}) and~(\ref{phiminus}) to the corresponding results in radiation (or matter) domination\footnote{We work out the matter domination case for completeness, but this may be a rather academic exercise since a very long dust epoch after inflation is hardly ever realised.}.  The latter can be found by solving the Bardeen equation in such backgrounds
\be
\Psi''+4\cH\Psi' = \frac{3\cH}{k^{7/2}} \left( \cH^2\ompip \right)' \, ,\\
\Psi''+3\cH\Psi' = \frac{3\cH}{k^{7/2}} \left( \cH^2\ompip \right)' \, ,
\ee
for radiation and matter domination, respectively, meaning that
\be
\Psi_{+}(x) &=& \Psi_0 + \frac{\Psi_1}{x^3} + \frac{3\ompip}{x^2} k^{-3/2} \, ,\label{psiplus}\\
\Phi_{+}(x) &=& \Psi_0 + \frac{\Psi_1}{x^3} \, , \label{phiplus}
\ee
in radiation domination, and
\be
\Psi_{+}(x) &=& \Psi_0 + \frac{\Psi_1}{x^5} + \frac{24\ompip}{x^2} k^{-3/2} \, ,\label{psiplusm}\\
\Phi_{+}(x) &=& \Psi_0 + \frac{\Psi_1}{x^5} + \frac{12\ompip}{x^2} k^{-3/2} \, , \label{phiplusm}
\ee
in matter domination; we have defined $\ompip \equiv \pisubs k^{3/2} / \rho_{\rm r,m}$ for radiation and matter, respectively -- note that in radiation domination $\ompip \propto \text{const}$ whereas in matter domination, since $\pisubs \propto 1/a^4$ and $\rho_{\rm m} \propto 1/a^3$ then $\ompip \propto 1/a$; useful relations are $\cH = 1/\eta$ in radiation domination, and $\cH = 2/\eta$ in matter domination ($\eta$ positive).

Notice the appearance of $\Psi_0$ and $\Psi_1$, two integration constants which have to be fixed by appropriate matching with the preceding inflationary epoch.

\subsection{Matching}

Since the energy-momentum tensor, in the idealised description we adopt, goes through a discontinuity when jumping out of inflation, special care must be taken in sewing the solutions of the Bardeen equations together.  Keeping in mind that we are working in the long wavelength limit, the appropriate way to ensure that the gravitational potentials are not discontinuous at the sharp transition is to keep the induced three-metric on and extrinsic curvature of the spacelike hypersurface which defines the transition continuous.  While this seems very pompous and technical, in the large-scale limit it simply reduces to demanding that the curvature perturbation
\be\label{xi}
\zeta \equiv \Psi + \frac{\cH^2}{\cH^2-\cH'} \left( \frac{\Psi'}{\cH} + \Phi \right) \, ,
\ee
is conserved through the transition.

The hypersurface which determines the transition is typically defined by the requirement that the energy density on such hypersurface remains constant; we will not dwell into the details of this procedure, and simply quote the result
\be\label{cT}
\cT = \frac{\delta\rho_\cph}{\rho_\cph'} \, ,
\ee
where $\cT$ defines the transition hypersurface.  Eq.~(\ref{cT}) refers to the transition from inflation to matter or radiation, but is applicable to other discontinuities of the same kind.  It is straightforward to write down the explicit form for $\cT$ in the matter or radiation cases, which looks like
\be\label{Trhofix}
\cT = \frac{1}{2\ep_1\cH} \left[ \frac{\rho_\eb}{\rho_\cph} + 2 \frac{\Psi_{-}'}{\cH} + 2 \Phi_{-} + \frac{2k^2}{3\cH^2} \Psi_{-} \right] \, ,
\ee
where all quantities have to be evaluated at the conformal time corresponding to the transition, which we call $\eta_\cT$.

In order to ensure that the Hubble parameter remains continuous through the transition we are automatically led to define the following, self-explanatory, relations:
\be\label{chofct}
\eta_{-} \equiv -\eta_\cT \equiv -\eta_{+} \quad {\rm or } \quad \eta_{-} \equiv -\eta_\cT/2 \equiv -\eta_{+}/2 \, ,
\ee
for radiation and matter domination, respectively.

With the definition of the transition hypersurface, Eq.~(\ref{cT}) all the necessary ingredients to finally pierce together the two halves of the cosmological evolution of the Bardeen potentials are in place; in the longitudinal gauge the matching conditions can be cast as
\be\label{matchlong}
[\cT]_\pm = 0 \, , \quad [\Psi]_\pm = 0 \, , \quad [\cH \Phi + \Psi' - (\cH^2 - \cH') \cT]_\pm = 0 \, .
\ee
Remembering that $\cH^2-\cH' = 2\cH^2$ and $\cH^2-\cH' = 3\cH^2/2$ for radiation and matter domination, respectively, one thus obtains the final expressions for the integration constant $\Psi_0$ (we are not interested in $\Psi_1$ since it decays rapidly):
\be
\Psi_0 &=& \frac{2}{3\ep_1} \left\{ b(k) + \cO(\ep^2) \frac{\ompi}{x_\cT^2} k^{-3/2} + \cO(\ep) \ompi k^{-3/2} + \left[ \frac{6-\alpha}{(3-\alpha)\alpha} + \frac{1}{2} \right] \omrho k^{-3/2} \right\} \, , \label{Psizero} \\
 \, ,
\ee
for ultra-relativistic gas, and similarly for dust.  The most important contribution comes from $\omrho$, while the similar $\ompi$ piece is suppressed by powers of the slow roll parameters; the largest term $\ompi/x_\cT^2$ is not present up to the next order in slow roll, but one can in fact show~\cite{Bonvin:2011dt} that it vanishes at all orders, as it should.

In principle one has to ensure that this perturbation does not grow disproportionately, or perturbation theory would be invalid.  Indeed, while, in this (dis)guise, the constant mode $\Psi_0$ seems rather innocuous, the contribution coming from the $\ompi$ term could in fact dramatically perilous.  An easy way to see this is by looking at the relative weight of this perturbed piece, which will enter the otherwise null Weyl tensor, with respect to the background, represented by the Ricci tensor:
\be\label{WoverR}
r^W_R \equiv \frac{\rm Weyl}{\rm Ricci} \simeq \frac{k^2\left(\Psi_{+}+\Phi_{+}\right)k^{3/2}}{\cH^2} \simeq x^2 \Omega_i(x_\cT) \, ;
\ee
it is obvious now that one has to normalise the overall anisotropic contribution in such a way that this ratio stays controllably below unity at all times, or at least until today.  This is automatically true as long as $x\ll1$ (large scales) and since we also need $\Omega_i(x_\cT)\ll1$ or the de Sitter background solution does not hold.  Notice that, had the $1/x_\cT^2$ mode leaked into $\Psi_0$, its contribution would be so much enhanced that, as it is easy to see, it would make magnetogenesis and inflation incompatible.

In any case, this new contribution to the constant mode is to be taken into account when inflationary predictions are discussed; in particular, it can lead to non-Gaussian signals, which we analyse in our specific model below.

\section{Three-form perturbations}\label{fatality}

The outcome of the previous section was that, in what seems to be a fairly general class of models of inflationary magnetogenesis, once one proceeds to analyse the first order gravitational perturbations induced by the presence of the magnetic (and electric) fields, there is a nonzero transmission coefficient for the magnetic energy density and anisotropic stress generated during inflation when moving over to radiation or matter domination.

We will now repeat the above calculation for the three-form case, pointing out where the key differences arise; the non-Gaussian signatures of this scenario will be discussed in the next section.

\subsection{Perturbations}

The first step is to write down a Bardeen equation for the three-form system.  In order to do so, we define the kinetic
\be\label{kp}
\quad K' \equiv X' + 3\cH X \, ;
\ee
the background equations for the three-form inflation system are
\be
&& X'' + 2\cH X' + 3\cH' X - 3\cH^2 X + a^2 V_{,X} = 0 \, ,\label{xeom}\\
&& \cH^2 = \frac{1}{3M_4^2} \left( \frac{1}{2} K'^2 + a^2 V_{,X} \right) \, ,\label{fx1}\\
&& \cH^2 - \cH' = \frac{1}{2M_4^2} a^2 V_{,X} X \, .\label{fx2}
\ee
In the longitudinal gauge the expressions for the perturbed Einstein tensor of course do not change (the three-form is minimally coupled); what does change are the formul\ae\ for the perturbed energy-momentum tensor.  The perturbation equations read
\be
&& 3\cH \Psi' + 3\cH^2 \Phi + k^2 \Psi = -4\pi G a^2 \delta\rho_X - 4\pi Ga^2 \rho_\eb \, , \label{psiprimeX}\\
&& \Psi''+ 2\cH \Psi' + \cH \Phi' + (2 \cH' + \cH^2) \Phi - \frac{k^2}{3} (\Phi-\Psi) = 4\pi G a^2 \delta P_X + 4\pi G a^2 p_\eb \, , \label{psidprimeX}\\
&& \Psi' + \cH \Phi = 4\pi G V_{,X} a\alpha - 4\pi G a \, {\rm i} \, \frac{k^j}{k^2}\, q_{\eb\, j} \, , \label{psiqX}\\
&& k^2 (\Phi-\Psi) = -8\pi G a^2 \pisubs \, . \label{phiminuspsiX}
\ee
The three-form perturbations are given by
\be\label{threepert}
\B_{0ij} \equiv a^3 \epsilon_{ijk} \left( B^T_k + \d_k B^L \right) \quad , \quad \B_{ijk} \equiv a^3 \epsilon_{ijk} \left( X + B_0 \right) \, ,
\ee
with $B^T_k$ the transverse (and ultimately only dynamical) degrees of freedom, and where
\be\label{eomXpert}
B_0' + 3\cH B_0 + \frac{V_{,X}}{X} a B^L + \frac{k^2}{a^2} a B^L + K' \left(3\Psi - \Phi\right) = 0 \, ,
\ee
and the energy and pressure density perturbations are given by
\be
a^2\delta\rho_X &=& K' \left[ B_0' + 3\cH B_0 + \frac{k^2}{a^2} a B^L + K' \left(3\Psi - \Phi\right) \right] + a^2 V_{,X} \left( B_0 + 3X\Psi \right) \, ,\label{drhoX}\\
a^2\delta P_X &=& -a^2\delta\rho_X + a^2 \left( V_{,XX} X + V_{,X} \right) \left( B_0 + 3X\Psi \right) \, .\label{dpX}
\ee

It is straightforward, if not extremely exciting, to combine these equations once more into a single second order differential equation for the Bardeen potentials, for instance $\Psi$:
\be\label{bardeenX}
\Psi'' - \frac{X'}{X} \left( 1 + \frac{V_{,XX} X}{V_{,X}} \right) \Psi' + \left[ 2\left(\cH' - \cH^2\right) - \cH \frac{X'}{X}\left( 1 + \frac{V_{,XX} X}{V_{,X}} \right) + \frac{V_{,XX} X}{V_{,X}} k^2 \right] \Psi = \source \, ,\nn\\
\ee
where the source now is
\be\label{sourceX}
\source = 8\pi G a^2 \!\!&&\!\! \left\{ \cH \frac{(a^2 \pisubs)'}{a^2 k^2} + \left[ 2\left(\cH' - \cH^2\right) - \cH \frac{X'}{X} \left( 1 + \frac{V_{,XX} X}{V_{,X}} \right) - \frac{k^2}{3} \right] \frac{\pisubs}{k^2} \right. \\
&& \!\!\!\! \left. - \frac{1}{2}(\frac{V_{,XX} X}{V_{,X}}\rho_\eb - p_\eb) + \frac{1}{2} \left( 3\cH - \frac{X'}{X} \right) \left( 1 + \frac{V_{,XX} X}{V_{,X}} \right) \frac{ {\rm i}\, k^jq_{\eb\, j}}{k^2 a} \right\} \, .
\ee
We can simplify this again for large scales using
\be\label{slowX}
\frac{X'}{X} \left( 1 + \frac{V_{,XX} X}{V_{,X}} \right) = 2\cH \left( \ep_1 + 3\ep_2 \right) \, ,
\ee
and the sound speed $c_S^2 \equiv V_{,XX} X / V_{,X}$ to obtain
\be\label{barEqX}
\frac{\dd^2 \Psi}{\dd x^2} + \frac{2(\ep_1 + 3\ep_2)}{x} \frac{\dd \Psi}{\dd x} + \bigg[c_S^2 - \frac{2(2\ep_1 + 3\ep_2)}{x^2} \bigg] \Psi = \cS_\eb \, ,
\ee
where the source takes almost the exact same form as in Eq.~(\ref{sourceLS}); the only relevant difference is thus the nontrivial speed of sound, which is in principle time-dependent (although only mildly so through the slow roll parameters):
\be\label{sourceXslowlyrolled}
\cS_\eb \simeq \frac{8\pi G a^2 \cH^2 }{k^4} \left[ \frac{\pisubs'}{\cH} + 2\left( 1 - 2\ep_1 - 3\ep_2 \right) \pisubs - \frac{k^2}{3} \pisubs - \frac{k^2}{2} (c_S^2 \rho_\eb - P_\eb) + \frac{3}{2}(c_S^2+1) k^2 Q_\eb \right] \, .
\ee

The solution to Eq.~(\ref{barEqX}) is again obtainable by resolving the homogeneous equation first, and then compose the particular inhomogeneous solution through the Wronskian method.  The homogeneous solutions are $\Psi_1(x) = x^p J_\nu(c_S x)$ and $\Psi_2(x) = x^p J_{-\nu}(c_S x)$, with, as before, $p=1/2+\mathcal{O}(\ep_1,\ep_2)$ and
$\nu=1/2+\mathcal{O}(\ep_1,\ep_2)$, and where we have taken the sound speed to be constant (up to slow roll corrections); thence, first of all, $\cW(x) = - \frac{2}{\pi} \sin(\nu\pi) x^{2p-1}$ as previously, which then means that the lowest order inhomogeneous solution will not know anything about the nontrivial speed of sound -- as is expected since this term is small at large scales:
\be
\label{psiinhX}
\Psi_{\rm inh}(\bk, x) &= \frac{\pi}{2\sin(\nu\pi)} \int_{x_{\rm h}}^x \dd \ln z \left(\frac{x}{z}\right)^p \!&\! \left[J_\nu(x) J_{-\nu}(z) - J_\nu(z) J_{-\nu}(x) \right] z^2 \cS_\eb(\bk, z) \nn\\
& \simeq \frac{1}{2\nu} \int_{x_{\rm h}}^x \dd \ln z \left(\frac{x}{z}\right)^p \!\!\!\!&\!\!\!\! \left\{\left(\frac{x}{z}\right)^\nu \left[ 1+\frac{c_S^2 z^2}{4\nu-4}-\frac{c_S^2 x^2}{4\nu+4} \right] \right. \\
&&\!\!\!\! - \left. \left(\frac{z}{x}\right)^\nu \left[ 1-\frac{c_S^2 z^2}{4\nu+4}+\frac{c_S^2 x^2}{4\nu-4} \right] \right\} z^2 \cS_\eb(\bk, z) \, . \nn
\ee

\subsection{Spectra}

While up to this point it looks like little has changed, one has in fact to remember that the convolution leading to the source term is very different from the $f^2F^2$ case -- recall we have $f=1$ in this example.  Indeed, the spectrum of the magnetic field is $P_B \propto \exp(2\Gamma)$, see~(\ref{cAdef1}), for horizon exit IC, and $P_B \propto \exp(2\Gamma k|\eta|)$ for IC at the beginning of inflation, Eq.~(\ref{cAdef2}).  The convolution in~(\ref{ppidef}) is an integral over momentum; due to the nontrivial $k$ dependence of the $\Gamma$ coefficient in the exponents, these integrals are very hard to do.  However, we know that the spectrum is essentially zero except for a narrow band of modes, where we can approximate $\Gamma$ as a simple power law in $k$.  In the simplest case, when $\Gamma=\text{const}$, the integral is doable analytically.  A difference to the scalar field case is that now there is also a contribution from the coupling term to the anisotropic stress in the stress energy tensor.  However, it is suppressed by the small coupling parameter and we neglect it here for simplicity.  Furthermore, noticing that $P_E = \Gamma P_{EB} \gg P_{EB} = \Gamma P_B \gg P_B$ we can approximate the convolution as
\be
P_\Pi &\simeq& \int \frac{\dd y y^3 |y-x|}{a^8 |\eta|^5} e^{2\Gamma(2-y-|x-y|))} \, , \quad \text{IC at } -1/k \, , \label{convX1}\\
P_\Pi &\simeq& \int \frac{\dd y y^3 |y-x|}{a^8 |\eta|^5} e^{2\Gamma(y-|y-x|)(1-\eta_i/\eta)} \, , \quad \text{IC at } \eta_i \, . \label{convX2}
\ee
The integrals have to be performed only over the range of modes which are amplified, that is, up to $k_\Lambda$ only.  Notice also that in the very end what we care about is the final time dependence, because this is what is going to tell us how to solve Eq.~(\ref{psiinhX}).

The most relevant difference here is that we need to explicitly account for the cutoffs, which we call $k_{IR}$ and $k_{UV}$, and the momentum integral will generally depend on these.  This will change significantly the shape of the result, for no longer the time dependence is going to be simplified at the boundaries $k=0$ and $k=1/|\eta|$.  Therefore,
\be
P_\Pi \simeq |\eta|^3 \cI(k, k_{IR}, k_{UV}) e^{2\Gamma(2-k\eta+2k_{uv}\eta)}\, , \quad \text{IC at } -1/k \, , \label{convX1s}\\
P_\Pi \simeq |\eta|^3 \cI(k, k_{IR}, k_{UV}) e^{2\Gamma(2k_{uv}-k)(\eta_i-\eta)}\, , \quad \text{IC at } \eta_i \, , \label{convX2s}
\ee
where in both cases the functions $\cI(k, k_{IR}, k_{UV})$ are polynomials whose shape depends on the specific values of the parameters $\Lambda$ and $k_\Lambda$; their form however does not matter for us since they do not depend on time explicitly -- moreover, if they are to solve the problem of today's observed large-scale magnetisation, they will return enough power at these lengthscales we are after.  Note that~(\ref{convX2s}) is valid as long as $\Gamma k_\Lambda |\eta| \ll 1$ -- were this not satisfied, the dominant contribution would be $\propto |\eta|^8$.

We can use these approximated expressions to calculate the form of the source in this case.  In the language of Eq.~(\ref{ppi0}) we would say that $\alpha = 3/2$ for the two above possibilities, and ignoring the mild time dependence coming from the exponential factors -- in computing $\pisubs'$ from~(\ref{eqpis}) this will simply add a subdominant term proportional to $\Gamma x$ to $(2-\alpha)$.

We can write down the solutions for the Bardeen potential up to terms of order $x^0$, which now contain information about the nontrivial speed of sound:
\be\label{psiminusX1}
\Psi_{-}(k) & \simeq & b(k) - \frac{3}{3-\alpha} \left[1+\frac{5-3\alpha}{3-\alpha}\ep_1 \right] \frac{\ompi}{x^2} k^{-3/2} \nn\\
&& - \frac{1}{(3-\alpha)(1-\alpha)} \left[ \ompi k^{-3/2} + \frac{2(6-\alpha)c_S^2 - (c_S^2-1)\alpha}{2\alpha} \omrho k^{-3/2} \right] \, ,
\ee
The potentially dangerous mode $\ompi/x^2$ again cancels away from the final expression for $\Psi_0$ just as it should; notice that since there is no sign of $c_S$ in the term proportional to the anisotropic stress $\ompi$, then once matched to radiation this contribution is going to be suppressed by slow roll parameters.

One can also look at this solution directly in terms of curvature $\zeta$, in which case this solution reads
\be\label{zetaX1}
\zeta_{\rm inh}(k) & \simeq & - \frac{3}{3-\alpha} \frac{1}{\ep_1} \left[ \ompi k^{-3/2} - \frac{2(6-\alpha)c_S^2 - (c_S^2-1)\alpha}{2\alpha} \omrho k^{-3/2} \right] \, .
\ee
Notice that this contribution is dominant with respect to the ``initial'' term when the integral is evaluated at $\eta_i$, for in the latter case the exponential factors become relevant and rapidly approach zero, killing the contribution (as it must be), whence the goodness of the approximated curvature~(\ref{zetaX1}).

\section{Non-Gaussianities}\label{nong}

The model we have developed introduces a direct coupling between the three-form and the one-form, that is, the vector potential.  Since in the final expressions for the curvature sourced by this coupling~(\ref{zetaX1}) the vector field always appears quadratically through its energy density $\rho_\eb$ or the anisotropic stress $\pisubs$, it is natural to expect some degree of non-Gaussianity in the primordial spectrum, even for purely Gaussian vector two-point functions.

Non-Gaussianity in the context of direct couplings between a scalar or pseudoscalar inflaton field and EM has been studied in~\cite{Barnaby:2011vw,Barnaby:2012tk}, from where we will borrow some of the machinery for our rough estimations.  The relevant difference from their work and ours is the form of our coupling term Eq.~(\ref{inte}), which does not resemble the structure of those most popular ones analysed in the literature.  
There will be some intrinsic non-Gaussianities arising from the third order coupled metric and three-form perturbations~\cite{Maldacena:2002vr,Seery:2005wm,Seery:2008ms}, but we expect this contribution to be subdominant, since the direct coupling is strongly enhanced by the rapid development of the vector potential instabilities (which in turn give rise to strong magnetic fields).

\subsection{The formalism}

Let us now sketch the steps, without repeating all the details which can be easily found in the literature, which lead to a reasonable guess of what level of non-Gaussianity we might reach in this context.  Since the calculations are, euphemistically, tedious, and in view of the approximations we are going to employ, we will limit ourselves to one significant example wherefrom illustrative conclusions can be drawn.

First of all, the leading contribution to the curvature comes from $\omrho$, so what we are interested in is the three-point function of this operator, that is $\langle\rho_\eb(\bp)\rho_\eb(\bq)\rho_\eb(\bk)\rangle$.  Each $\rho_\eb$ at the operator level will contain combinations of $|\cA(k)'|^2$, $|k\cA(k)'\cA(k)|$ and $k^2|\cA(k)|^2$; we have seen that the parameter space which is relevant for magnetogenesis is that for which, schematically: $\cA' \simeq \Gamma k \cA \gtrsim k \cA$ since $\Gamma \gtrsim 1$.  Thence, we only retain the electric field components in $T_{\mu\nu}$.

The single $a^4\rho_\eb(\bk,\eta)$ will be a convolution of the type
\be\label{rhoconv}
\int\! \dd^3k'\sum_{\lambda\,\lambda'} \left\{ \cep_i^\lambda(\bk') \cep^{i\lambda'}(|\bk+\bk'|) a_\lambda(\bk') a_{\lambda'}(|\bk+\bk'|) k'|\bk+\bk'| \cA'(k')\cA'(|\bk+\bk'|) + 3\times{\rm h.c.} \right\} \, , \nn\\
\ee
where by $3\times{\rm h.c.}$ we mean the remaining three combinations of creation and annihilation operators\footnote{Notice that here we are not taking the expectation value, as it is not a fluctuation, but we retain the full vectorial form and bra-ket it only at the end as usual to capture the interferences.}.  The three-point function is going to contain all possible combinations of three such terms; let us look at just one of them, that with structure $\langle a a a^\dagger a a^\dagger a^\dagger \rangle$:
\be\label{aaaaaa}
\langle \rho_\eb^3 \rangle & \supset & \frac{1}{8a^{12}} \int \frac{\dd^3k'\dd^3p'\dd^3q'}{(2\pi)^{9/2}} \sum_{\lambda\,\lambda'} \sum_{\mu\,\mu'} \sum_{\nu\,\nu'} \Big\{ \nn\\
&& \cep_i^\lambda(\bk') \cep^{i\lambda'}(|\bk+\bk'|) \cep_j^{\mu\,*}(\bp') \cep^{j\mu'}(|\bp'-\bp|) \cep_k^{\nu\,*}(\bq') \cep^{k\nu'\,*}(|\bq-\bq'|) \nn\\
&& \langle a_\lambda(\bk') a_{\lambda'}(|\bk+\bk'|) a_\mu(\bp') a^\dagger_{\mu'}(|\bp'-\bp|) a^\dagger_\nu(\bq') a^\dagger_{\nu'}(|\bq-\bq'|) \rangle \nn\\
&& k'|\bk+\bk'| p'|\bp'-\bp| q'|\bq-\bq'| \Gamma(k') \Gamma(|\bk+\bk'|) \Gamma(p') \Gamma(|\bp'-\bp|) \Gamma(q') \Gamma(|\bq-\bq'|) \nn\\
&& \cA(k')\cA(|\bk+\bk'|) \cA^*(p')\cA(|\bp'-\bp|) \cA^*(q')\cA^*(|\bq-\bq'|) \Big\} \, ,
\ee
and a number of other similar terms.  We chose this one because it is one nontrivial one since it it not simply a product of $\langle \rho_\eb \rangle \langle \rho_\eb^2 \rangle$ which would be uninteresting.

The expectation value $\langle a a a^\dagger a a^\dagger a^\dagger \rangle$ returns a series of four combinations of, schematically $(\delta_{\rho\sigma})^3 \times (\delta^3(\br-\bs))^3$ which will lead all to very similar terms; we pick the representative one
\be\label{aaaaaaVEV}
\delta_{\lambda\mu}\delta_{\mu'\nu}\delta_{\lambda'\nu'} \delta^3(\bp'-\bk')\delta^3(\bp'-\bp-\bq')\delta^3(\bk+\bk'+\bq-\bq') \, .
\ee
This will take care of three summations (over $\lambda'$, $\mu'$, and $\nu'$) and two integrals, leaving us with one convolution only; the third $\delta^3$ in all terms is nothing else than $\delta^3(\bp+\bq+\bk)$ which enforces the ``triangularity'' of the bispectrum, descending from translation invariance.  Finally, we deal with the polarisation vectors $\cep$ with the aid of the formula
\be\label{cepsum}
\sum_{\alpha\beta} \cep_i^\alpha(\bk) \cep^{i\beta}(\bk') = 1 + \left(\hat k\cdot\hat k' \right)^2 \, ,
\ee
which directly leads to
\be\label{fullI}
\langle \rho_\eb^3 \rangle & \supset & \frac{1}{8a^{12}} \int \frac{\dd^3k'}{(2\pi)^{9/2}} k'|\bk'+\bk| k'|\bk'-\bp| |\bk'-\bp||\bk'-\bp-\bq| \delta^3(\bp+\bq+\bk) \nn\\
&& \Gamma(k') \Gamma(|\bk+\bk'|) \Gamma(k') \Gamma(|\bk'-\bp|) \Gamma(|\bk'-\bp|) \Gamma(|\bk'-\bp-\bq|) \nn\\
&& \cA(k')\cA(|\bk+\bk'|) \cA(k')\cA(|\bk'-\bp|) \cA(|\bk'-\bp|)\cA(|\bk'-\bp-\bq|) \nn\\
&& \left\{ \left[ 1 + \left( 1 + \hat k \hat k' \right)^2 \right] \left[ 1 + \left( 1 - \hat p \hat k' \right)^2 \right] \left[ 1 + \left( 2 - 2\hat p \hat k' - \hat q \hat k' + \hat p \hat q \right)^2 \right] \right\} \, ,
\ee
where we know that our solutions $\cA(k)$ are real-valued.

Some approximations are necessary at this point.  We focus on one of the two cases examined in Sec.~\ref{threeEM}, that is, the case where the IC are fixed at $\eta = -1/k$, which means that $\sqrt{2k} \cA \simeq \exp \Gamma$ where we neglect its (subdominant for large scales) time dependence.  In order to facilitate the numerical evaluation of~(\ref{fullI}) we also employ the same simplification we adopted for the study of magnetogenesis, and consider $\Gamma(k) \simeq 1/\Lambda$ with a cutoff at $k \approx \Lambda k_\Lambda$, from which point on $\Gamma$ decays, and eventually becomes imaginary at $k \rar k_\Lambda$.  It is a very rough approximation, but given the extreme sensibility that we expect from the parameters $\Lambda$ and $k_\Lambda$ in view of the exponential factor, it will serve the scope of this exposition.  Collecting these choices leads at once to
\be\label{simpleI}
\langle \rho_\eb^3 \rangle & \supset & \frac{1}{64a^{12}} e^{6\Gamma} \Gamma^6 \int \frac{\dd^3k'}{(2\pi)^{9/2}} |\bk'+\bk| |\bk'-\bp| |\bk'-\bp-\bq| \, \delta^3(\bp+\bq+\bk) \nn\\
&& \left\{ \left[ 1 + \left( 1 + \hat k \hat k' \right)^2 \right] \left[ 1 + \left( 1 - \hat p \hat k' \right)^2 \right] \left[ 1 + \left( 2 - 2\hat p \hat k' - \hat q \hat k' + \hat p \hat q \right)^2 \right] \right\} \, .
\ee

We study this integral numerically; first of all, we employ the delta function to fix $\bk = -(\bp+\bq)$, which eliminates three out of nine variables (excluding the free parameters $\Lambda$ and $k_\Lambda$), and then choose a convenient reference frame -- this does not affect the generality of the outcome -- as
\be\label{frame}
\bp = (0,0,p) \,,\quad \bq = (0,q_2,q_3) \,,\quad \bk = (0,-q_2,-q_3-p) \, .
\ee
At this point since there still are three variables ($q_2$, $q_3$, and $p$), one needs to specify a shape for the triangle under inspection.  This in general will not guarantee that the resulting non-Gaussianity follows this pattern, but one can assess the goodness of the template one tampers with.  In our case we are essentially interested in looking for possible new shapes, and more than anything else we are keen to see whether our model can at once provide magnetogenesis and detectable non-Gaussianities.

One possible way to characterise the level of non-Gaussian signal is of course to look for the deviations from simple Gaussian power spectra in the form of its variance.  This leads to the definition of the $f_{\rm NL}$ parameter for the curvature perturbations $\zeta$ through its spectrum as
\be
\langle \zeta(\bp) \zeta(\bq) \rangle & \equiv & \frac{2\pi^2}{k^3} P_\zeta(k) \delta^3(\bp+\bq) \label{Pzeta} \, , \\
\langle \zeta(\bp) \zeta(\bq) \zeta(\bk) \rangle & \equiv & \frac{3}{10} (2\pi)^{5/2} f_{\rm NL} P_\zeta^2(k) \delta^3(\bp+\bq+\bk) \frac{p^3+q^3+k^3}{p^3q^3k^3} \label{P2zeta} \\
	& \equiv & B_\zeta \delta^3(\bp+\bq+\bk) \label{Bzeta} \, .
\ee
Looking at our solution for $\zeta$ Eq.~(\ref{zetaX1}) we can infer $\langle \zeta \rangle \simeq c(\alpha) \langle \rho_\eb \rangle / \ep_1 \rho_\cph$ and obtain
\be\label{fnl}
f_{\rm NL} \simeq \frac{15}{64(2\pi)^3\ep_1} \left[ c(\alpha)e^{2\Gamma}\Gamma^2 \right]^3 \left[\frac{\left(\Lambda k_\Lambda\right)^4}{a_e^4\rho_X}\right]^3 \left( \frac{M_\text{P}^4}{\rho_X} \right)^2 \mathscr{I}(\bp,\bq,\bk) \,.
\ee
where we have factorised all shape-related factors in the term
\be\label{fnlI}
\mathscr{I}(\bp,\bq,\bk) & \equiv & \frac{p^3+q^3+k^3}{p^3q^3k^3} \int \dd^3k' |\bk'+\bk| |\bk'-\bp| |\bk'-\bp-\bq| \\
&& \left\{ \left[ 1 + \left( 1 + \hat k \hat k' \right)^2 \right] \left[ 1 + \left( 1 - \hat p \hat k' \right)^2 \right] \left[ 1 + \left( 2 - 2\hat p \hat k' - \hat q \hat k' + \hat p \hat q \right)^2 \right] \right\} \, ;\nn
\ee
notice that all momenta appearing in the last Eq.~(\ref{fnlI}) are normalised to $\Lambda k_\Lambda$, and therefore run from 0 to 1.  This integral can be fed to numerical solvers once a shape is specified, and is expected to not run amok: the all-important amplitude factors are displayed in the prefactor.

A comment is in order here.  In performing the integration we have approximated $\Gamma$ as a constant in momentum, and cut this at $\Lambda k_\Lambda$ because from that point on the actual value of $\Gamma$ decays as $1/k^2$ and is eventually cut off completely shortly thereafter (at $k = k_\Lambda$ itself).  This implies that, since we have six $\Gamma$s in the integrand, looking at their arguments we see that all the momenta involved (the internal $\bk'$ as well as the external $\bp$, $\bq$, and $\bk$) will need to be limited by the same threshold.  This implies that, just as it happens for the magnetic field which is only generated at the largest scales -- for low energy cut off momentum, so the non-Gaussianity produced in this model will be limited to large-scale correlations.  This is a new feature which will help discerning this model from the plenitude of other variants available in the literature.

\subsection{Results}



\begin{figure}[ht]
\begin{center}
\includegraphics[width=0.45\columnwidth]{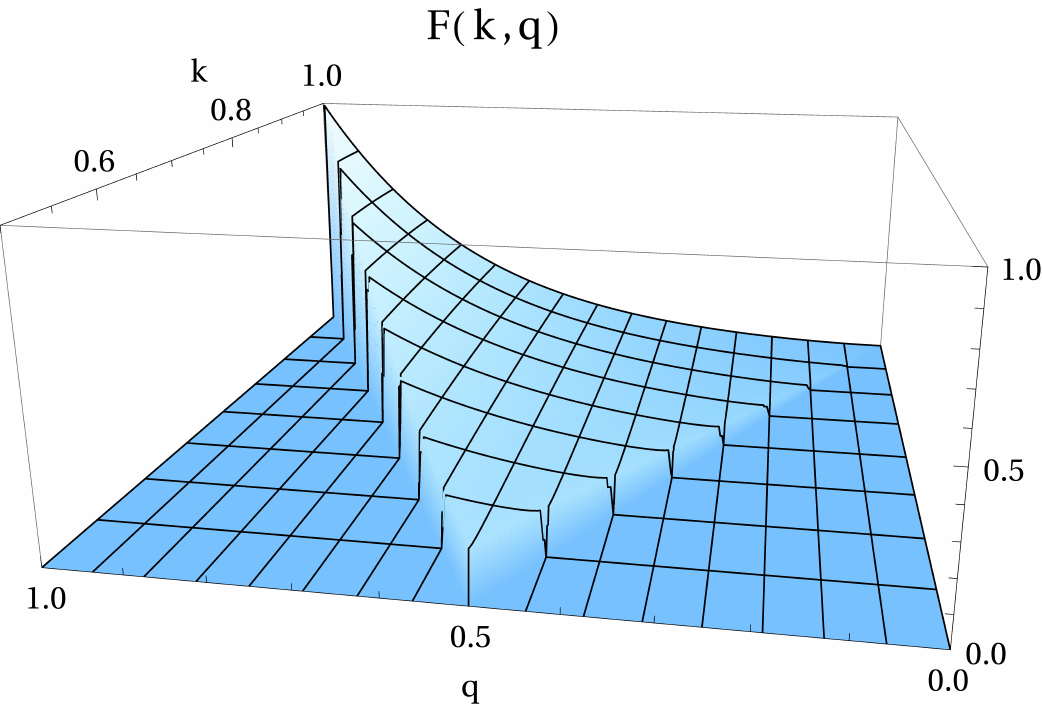} \label{shapes}
\includegraphics[width=0.45\columnwidth]{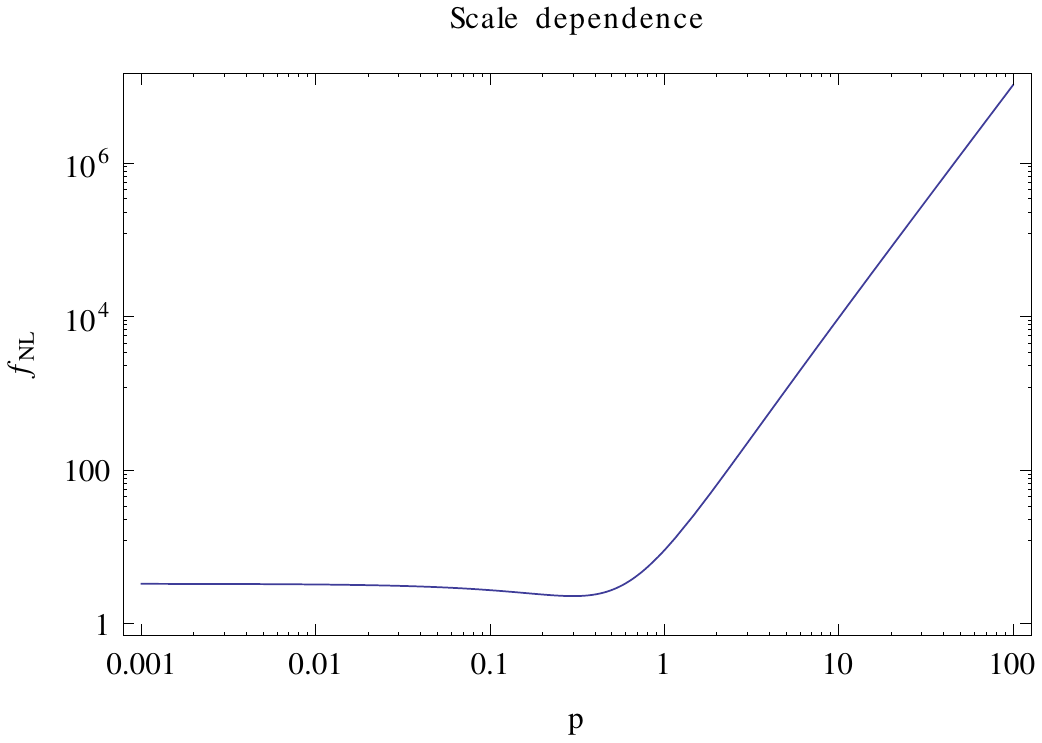} 
\caption{\emph{Right}.  The non-Gaussian shape function $F(q,k)$ defined in the text.  In our case it clearly peaks at the equilateral configuration $q=k=p$.  \emph{Left}.  The scale dependence for equilateral non-Gaussianity.}
\end{center}
\end{figure}

To first determine the shape of the non-Gaussianities at hand, we employ the visualisation introduced in~\cite{Babich:2004gb}.  Then we define the quantity 
\be\label{zeta3}
\langle \zeta_{\bf q} \zeta_{\bf k} \zeta_{\bf p} \rangle = (2\pi)^2 F({\bf q},{\bf k}, {\bf p}) \, . 
\ee
Without loss of generality, we order the momenta as $q \le k \le p$.  The relevant function is then $F(q/p,k/p)$, which is normalised to unity for the equilateral configuration: $F(1,1)=1$.  This function quantifies the relative contribution of different non-Gaussian shapes to the observable CMB anisotropies.  To avoid plotting the same region twice, we plot only the region $q<k$, $q>1-k$.  The result is shown in Figure~\ref{shapes}.

We see that the non-Gaussianity peaks at the equilateral shapes.  This is what we might have expected, since equilaterality is the typical outcome for perturbations generated within the horizon: in the present model we have seen that since the exponential terms, responsible for the perturbation generation, blow up quickly at the beginning after we set the IC and then become constant.  Having verified this numerically, we will restrict to equilateral configurations in the following and set $q=k=p$.

In order to grasp what the scale dependence is going to be, we plot the an arbitrarily normalised non-Gaussianity in the left panel of figure~\ref{shapes}.  For large scales, $f_{NL}$ remains a small constant, but for the subhorizon scales there is a strong amplification with the increasing wavemode.  The numerical value of non-Gaussianity will depend upon the model parameters, and in the rest of this section we study the compatibility of the model parameters with the two-fold requirement of both sufficient magnetic field generation and non-Gaussianity that lies in the interesting range of roughly $1<|f_{NL}|<100$.  

In the specific case under scrutiny said parameters are the UV cutoff scale $k_\Lambda$ and the strength of the coupling $\Lambda$.  As is expected (recall the study of magnetogenesis in Sec.~\ref{threeEM} the result will be exponentially sensitive to variations in $\Lambda$; in waiting for the results of the full numerical integration (also with the numerical solution of the Euler-Lagrange equation~(\ref{eom_a3}) -- recall we are working with an idealised de Sitter space) what we want to show here is that it is \emph{possible} to generate large-scale magnetic fields \emph{and}, at the same time, realise a detectable beyond-Gaussian imprint in the CMB.

\begin{figure}[ht]
\centering
\includegraphics[width=0.45\textwidth]{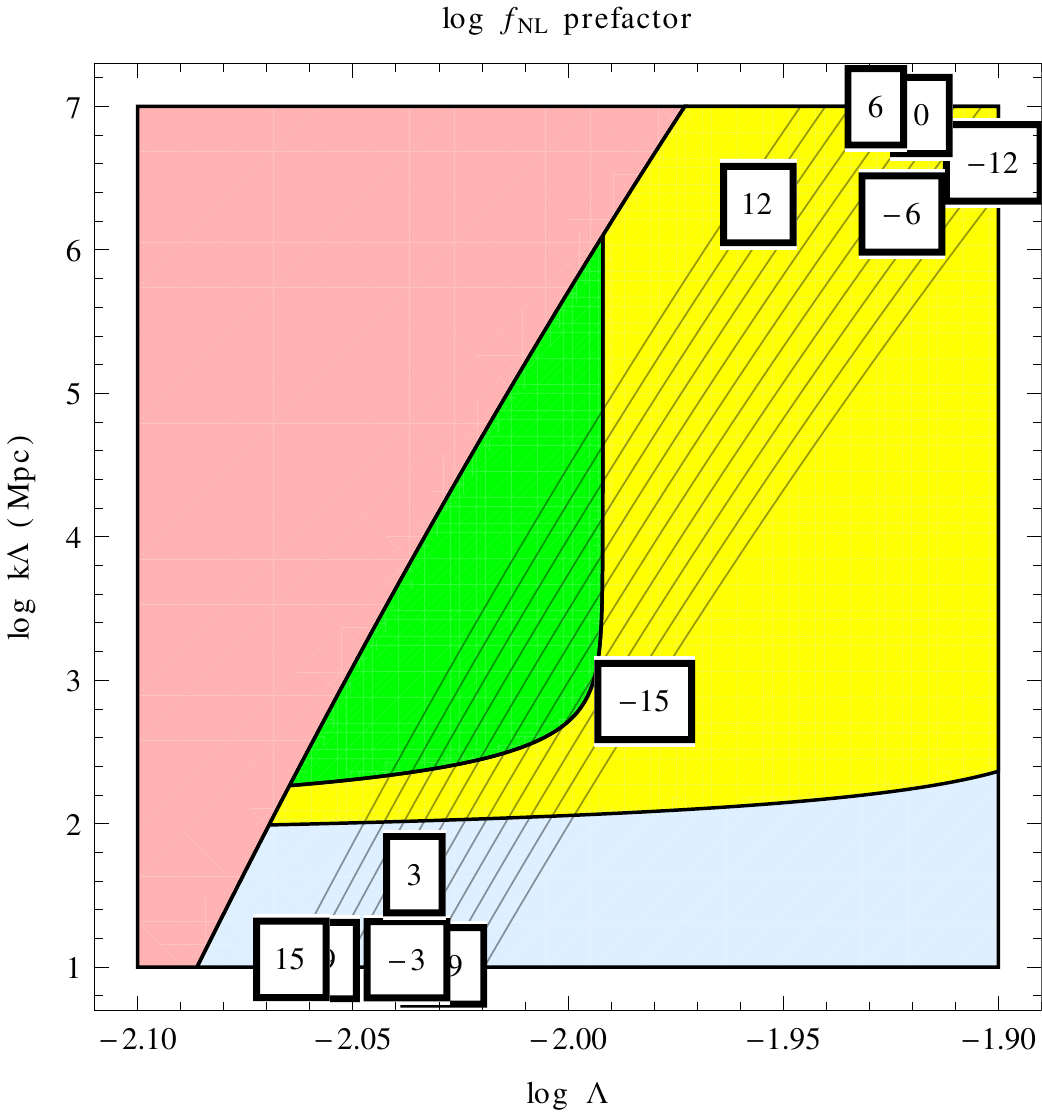}
\includegraphics[width=0.45\textwidth]{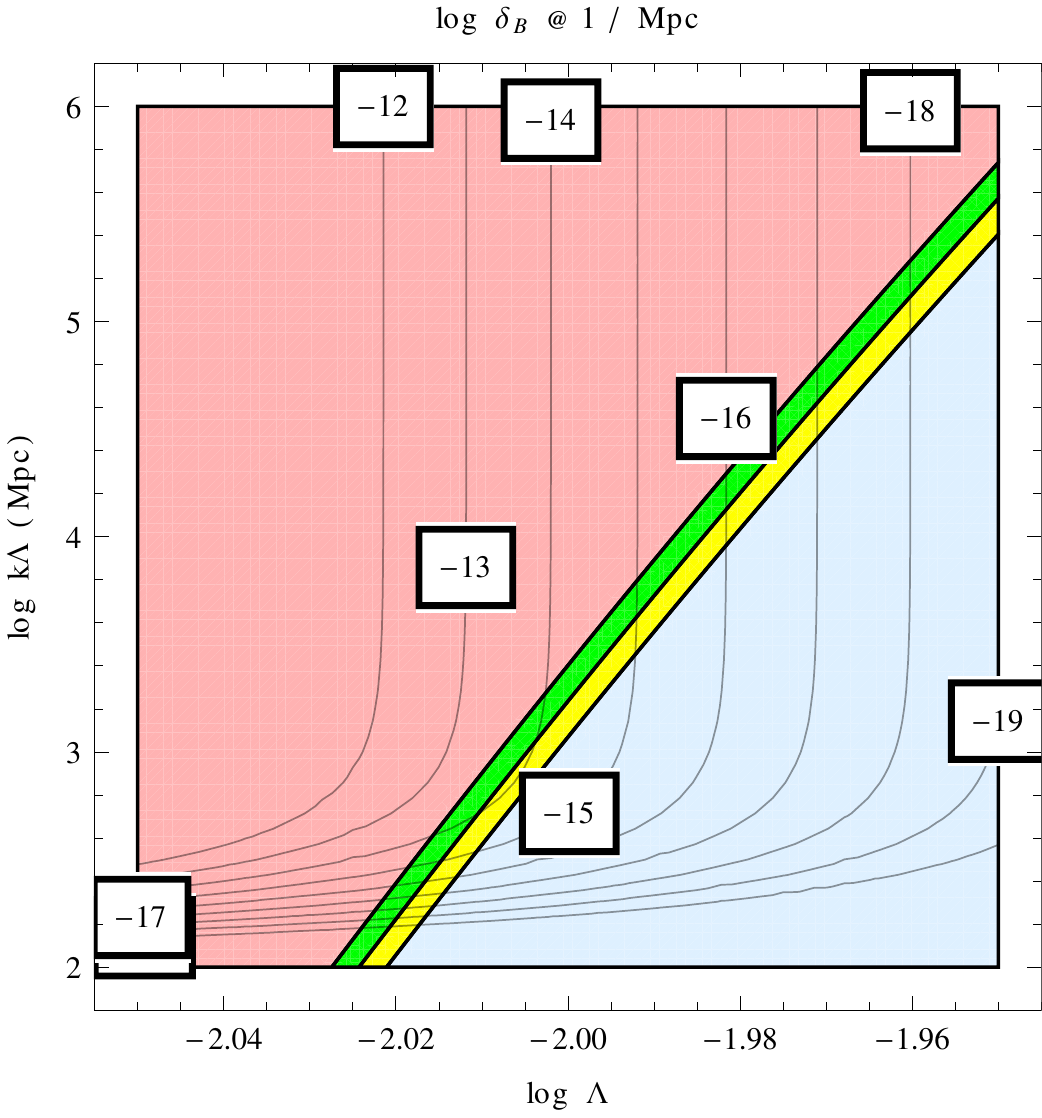}
\caption{\emph{Left}.  We zoom onto the most interesting region of parameter space of our Fig.~\ref{fig1} (same colours), while we superimpose the values of the prefactor to $f_{\rm NL}$ from Eq.~(\ref{fnlI}) in a logarithmic scale.  \emph{Right}.  Here make contour plots for the regions where this coefficient falls in the ranges: smaller than $10^{-2}$ (light blue, uninteresting), $10^{-2}$ to $1$ (yellow, small), $1$ to $100$ (green, detectable), and beyond $100$ (light red, dangerous); on top of these regions are drawn the curves of constant magnetic field strength at 1/Mpc (in Gauss units).}
\label{fig3}
\end{figure}

In Fig~\ref{fig3}, left, we zoom onto the most interesting region of the $(\Lambda,k_\Lambda)$ parameter space, that where the generated magnetic fields could match those required by observations on large scales.  The light blue region is that for which the field strength is below $10^{-25}$ Gauss at 1/Mpc, and is uninteresting phenomenologically; the yellow region goes up to $10^{-15}$ Gauss, and would, with the help of some dynamo mechanism, make up for observations; the green region has a field stronger than $10^{-15}$ and can directly account for Mpc scale magnetisation; the light red region is excluded due to background backreactions constraints.  The coefficient for $f_{\rm NL}$ which can be read off Eq~(\ref{fnlI}) can vary significantly in this region, due to the exponential sensitivity to the coupling and the high power with which the cutoff scale appears; this coefficient then will be enhanced (suppressed) by up (down) to a couple of orders of magnitude when the full integration and momentum dependence is considered, but it still is useful to understand the basic picture.  We see how it is possible to obtain order one non-Gaussianities in the green and yellow areas: this was highly nontrivial in the beginning; we plan to explore the details of such signatures (shapes, different potentials, realistic inflationary set-up, more detailed backreaction bounds, tensor perturbations) in a forthcoming work.

The right hand side of Fig~\ref{fig3} instead shows the regions where the $f_{\rm NL}$ coefficient is below 0.01 (light blue, deemed uninteresting), between 0.01 and up to 1 (yellow, small non-Gaussianity), between 1 and 100 (green, detectable non-Gaussianity), and beyond 100 (light red, too much power in the bispectrum).  The yellow and green bands are going to be shifted to the left or right, but only by some very small amount since we do not expect the integral $\mathscr{I}$ to be much different from one.  On top of these regions we show the values for the generated magnetic field at the Mpc scale.  This is in some sense the inverse or complementary plot to the left hand side one, and once more demonstrates how we can find some bands where the non-Gaussian signal is strong enough to be measurable, and at the same time large scale magnetisation is successfully achieved.

What we read off these figures is the effect of the natural cutoff of three-form inflation.  In the first place this affords to amplify extremely rapidly the EM field once the univocal $U(1)$-invariant coupling is introduced, without incurring in the most basic backreaction effect of blocking de Sitter expansion.  Since the scale $k_\Lambda$ can be chosen to lie in the low-energy range, the band of magnetic field which is boosted is very narrow; the spectrum also decays in the IR, ensuring that no divergence is hiding there for any duration of inflation.

Secondly, the UV threshold limits from above the size of the non-Gaussian scale which is going to be processed by the coupling.  This limit automatically gives birth to interesting phenomenology, shapes that are not obvious, and peculiar running with the scale, mimicking that of the generated EM field.  It is instructive to compare our findings with those of~\cite{Barnaby:2012tk}: in their case the logarithmic enhancement of the bispectrum is due to a very long period of inflation before the CMB scale has exited the horizon, and is a result of the vector waves being amplified from very low energies (largest scales) which all contribute to the bispectrum since their spectrum is flat.  In our scenario on the other hand, for $k_\Lambda$ not tremendously larger than the Hubble scale today, everything happens only at the scales of interest.  In the end it seems possible to reconcile magnetogenesis and its epiphenomenal non-Gaussianity.

\section{Summary}\label{end}

Three-form inflation has proven to be an enticing arena for the phenomenology of EM couplings to inflation.  We have analysed in more detail the unique gauge-invariant coupling which is allowed between the three-form and the EM one-form, going beyond our previous background findings.  The outcomes are bold and encouraging us to undertake further work in this direction.

First of all, ``three-magnetogenesis'' proves to be perhaps the only one magnetogenesis model which eschews strong coupling and background backreaction bounds, yet is capable of amplifying EM fluctuations on large scales to a degree which would be compatible with empirical evidence.

Secondly, at the first-order perturbation level one finds, in agreement with recent works, that the constant mode of the Bardeen potential and the curvature perturbations after inflation ends are influenced directly by the spectrum of the produced vector fluctuations on large scales.  In the simplest case we have described such fluctuations are statistically isotropic, but one can easily imagine extensions to more general cases.

In the third place, the ``revised'' curvature perturbation, containing bilinears in the Gaussian EM spectra, is genuinely non-Gaussian.  We have explored in some detail the simplest model to find captivating hints to nontrivial new signatures, shapes, and amplitudes within the detectable span.  What we found was scale-dependent equilateral non-Gaussianity that is within the viable and future-observable range, where at the same time the desired amount of magnetic field is generated.  Even though more precise predictions are demanded to be of use for data analysis, the toy-model we have adopted for the latter part of this study already contains all the important physics, in particular the central r\^ole played by the three-form duality, which translates in a low-energy UV cutoff of the enhanced vector spectra.

This model's features permit a direct link between magnetogenesis, inflation, and bispectra, in a way that can be reconciled with multiple observational pieces of evidence, a task definitely not paltry; the positive results of our analysis prompt us to go beyond the na\:ive analytical estimations we have presented, in order to fully eviscerate the precise phenomenology of this model.

\section*{Acknowledgements}
TK is supported by the Research Council of Norway.  FU thanks Chiara Caprini for hospitality at the IPhT, Saclay, and valuable insight; he is supported by the Research Council of Norway through the Leiv Eiriksson mobility programme fellowship N.216785/F11.

\bibliography{3fBfatal}
\bibliographystyle{JHEP}

\end{document}